# An enthalpy-based multiple-relaxation-time lattice Boltzmann method for solid-liquid phase change heat transfer in metal foams

Qing Liu,[1] Ya-Ling He,[1] and Qing Li[2]

[1]Key Laboratory of Thermo-Fluid Science and Engineering of Ministry of Education, School of Energy and Power Engineering, Xi'an Jiaotong University, Xi'an, Shaanxi, 710049, China

[2]School of Energy Science and Engineering, Central South University, Changsha 410083, China

In this paper, an enthalpy-based multiple-relaxation-time (MRT) lattice Boltzmann (LB) method is developed for solid-liquid phase change heat transfer in metal foams under local thermal non-equilibrium (LTNE) condition. The enthalpy-based MRT-LB method consists of three different MRT-LB models: one for flow field based on the generalized non-Darcy model, and the other two for phase change material (PCM) and metal foam temperature fields described by the LTNE model. The moving solid-liquid phase interface is implicitly tracked through the liquid fraction, which is simultaneously obtained when the energy equations of PCM and metal foam are solved. The present method has several distinctive features. First, as compared with previous studies, the present method avoids the iteration procedure, thus it retains the inherent merits of the standard LB method and is superior over the iteration method in terms of accuracy and computational efficiency. Second, a volumetric LB scheme instead of the bounce-back scheme is employed to realize the no-slip velocity condition in the interface and solid phase regions, which is consistent with the actual situation. Last but not least, the MRT collision model is employed, and with additional degrees of freedom, it has the ability to reduce the numerical diffusion across phase interface induced by solid-liquid phase change. Numerical tests demonstrate that the present method can be served as an accurate and efficient numerical tool for studying metal foam enhanced solid-liquid phase change heat transfer in latent heat storage. Finally, comparisons and discussions are made to offer useful information for practical applications of the present method.

## I. INTRODUCTION

Over the past three decades, latent heat storage (LHS) using solid-liquid phase change materials (PCMs) has attracted a great deal of attention because it is of great importance for energy saving, efficient and rational utilization of available resources, and optimum utilization of renewable energies [1-5]. Solid-liquid PCMs absorb or release thermal energy by taking advantage of their latent heat (heat of fusion) during solid to liquid or liquid to solid phase change process. PCMs have many desirable properties, such as high energy storage density, nearly constant phase change temperature, small volume change, etc. However, the available PCMs commonly suffer from low thermal conductivities (in the range of 0.2~0.6 W/(m·K) [1]), which prolong the thermal energy charging and discharging period. In order to overcome this limitation and improve the thermal performance of LHS units/systems, a lot of heat transfer enhancement approaches have been developed, among which embedding PCMs in highly conductive porous materials (e.g., metal foams, expanded graphite) to form composite phase change materials (CPCMs) has long been practiced [6]. High porosity open-cell metal foams, as a kind of promising porous materials with high thermal conductivity, large specific surface area, and attractive stiffness/strength properties, have been widely used for LHS applications [7].

With new experimental techniques and advanced instruments, experimental investigations of heat transfer behaviors in porous systems are becoming more accessible, and the problems of solid-liquid phase change heat transfer in metal-foam-based PCMs have been experimentally studied by many researchers [8-12]. In addition to experimental studies, numerical analyses usually play an important role in studying such problems. In the past two decades, numerical investigations have been extensively conducted to study solid-liquid phase change heat transfer in metal foams [13-23]. These numerical investigations provide valuable design guidelines for practical applications of LHS technologies. Since the thermal conductivity of the metal foam is usually two or three orders of magnitude higher than that of the PCM, the thermal non-equilibrium effects between the PCM and metal foam may play a significant role. Therefore, the local thermal non-equilibrium (LTNE) model (also called the two-temperature model) has been widely employed for numerical studies [14-23]. However, most of the previous numerical studies [13-20] for solid-liquid phase change heat transfer in metal foams were carried out using conventional numerical methods [mainly finite-volume method (FVM)] based on the discretization of the macroscopic continuum equations. In order to get a thorough understanding of the underlying mechanisms, more fundamental approaches should be developed for solid-liquid phase change heat transfer in metal foams.

The lattice Boltzmann (LB) method [24-28], as a mesoscopic numerical method sitting in the intermediate region between microscopic molecular dynamics (MD) and macroscopic continuum-based methods, has achieved great success in simulating fluid flows and modeling physics in fluids since its emergence in 1988 [29-33]. Historically, the LB method originated from the lattice gas automata (LGA) method [34], a simplified, fictitious version of the MD method in which the time, space, and particle velocities are all discrete. Later He and Luo [35,36] demonstrated that the LB equation can be rigorously obtained from the linearized continuous Boltzmann equation of the single-particle distribution function. The establishment of such connection not only makes the LB method more amenable to numerical analysis, but also puts the LB method on the solid theoretical foundation of kinetic theory. From this perspective, the LB method can be viewed as a Boltzmann equation-based mesoscopic method. Between the microscopic MD and macroscopic continuum-based methods, there also exist several other Boltzmann equation-based mesoscopic methods, such as the discrete-velocity method (DVM) [37] and the gas-kinetic scheme (GKS) [38,39], as representatives. Unlike the MD method which takes into account the movements and collisions of all the individual molecules, the LB method considers the behaviors of a collection of pseudo-particles (a pseudo-particle is comprised of a large number of molecules) moving on a regular lattice with particles residing on the nodes. This feature of the LB method is similar to that of the direct simulation Monte Carlo (DSMC) method [40-42]. Different from the conventional numerical methods based on a direct discretization of the macroscopic continuum equations, the LB method is based on minimal lattice formulations of the continuous Boltzmann equation for single-particle distribution function, and macroscopic properties can be obtained from the distribution function through moment integrations. As highlighted by Succi [43], the LB method should most appropriately be considered not just as a smart Navier-Stokes solver in disguise, but rather like a fully-fledged modeling strategy for a wide range of complex phenomena and processes across scales.

In recent years, the LB method in conjunction with the enthalpy method has been successfully employed to simulate solid-liquid phase change heat transfer in metal foams [21-23]. Gao et al. [21] proposed a thermal LB model to simulate melting process coupled with natural convection in open-cell metal foams under LTNE condition. The influence of foam porosity and pore size on the melting process were investigated and discussed. Subsequently, Gao et al. [22] further developed a thermal LB model for solid-liquid phase change in metal foams under LTNE condition. By appropriately choosing the equilibrium temperature distribution functions and discrete source terms, the energy equations of the PCM and metal foam can be exactly recovered. Most recently, Tao et al. [23] employed an enthalpy-based LB method to study the LHS performance of copper foams/paraffin CPCM. The effects of geometric parameters such as pore density and porosity on PCM melting rate, thermal energy storage capacity and density were investigated.

Up to now, although some progresses have been made in studying solid-liquid phase change heat transfer in metal foams, there are still two key issues remain to be resolved. The first one is to avoid iteration procedure so as to improve the accuracy and computational efficiency. In previous studies [21-23], the nonlinear latent heat source term accounting for the phase change is treated as a source term in the LB equation of the PCM temperature field, which makes the explicit time-matching LB equation to be implicit. Therefore, an additional iteration procedure is needed at each time step so that the convergent solution of the implicit LB equation can be obtained, which severely affects the computational efficiency, and the inherent merits of the LB method are lost. The second key issue is to accurately realize the no-slip velocity condition in the interface and solid phase regions. For solid-liquid phase change heat transfer in metal foams, the phase interface is actually a region with a certain thickness because of the interfacial heat transfer between PCM and metal foam [15]. Therefore, the phase interface is usually referred as the interface region or mushy zone. Considering the actual situation of the phase change process, it is not appropriate to use the bounce-back scheme to impose the no-slip velocity condition in the interface region (this point will be demonstrated in Section V B).

In the present study, we aim to develop a novel enthalpy-based LB method for solid-liquid phase change heat transfer in metal foams, in which the above-mentioned key issues will be resolved. Considering that the multiple-relaxation-time (MRT) collision model [28] is superior over its

Bhatnagar-Gross-Krook (BGK) counterpart [27] in simulating solid-liquid phase change heat transfer in metal foams, the MRT collision model is employed in the enthalpy-based LB method. We will compare these two collision models in Section V A. The rest of this paper is organized as follows. The macroscopic governing equations are briefly given in Section II. Section III presents the enthalpy-based MRT-LB method in detail. Section IV validates the enthalpy-based MRT-LB method. In Section V, comparisons and discussions are made to offer useful information for practical applications of the present method. Finally, some conclusions are given in Section VI.

## II. MACROSCOPIC GOVERNING EQUATIONS

For solid-liquid phase change heat transfer coupled with natural convection in metal foams, the following assumptions are made: (1) the flow is incompressible and laminar; (2) the thermophysical properties of the metal foam ($m$) and PCM ($f$) are constant over the range of temperatures considered, but may be different for the metal foam, liquid PCM ($l$) and solid PCM ($s$); (3) the metal foam and PCM are homogeneous and isotropic, the metal foam/solid PCM are rigid, and the porosity of the metal foam is constant; (4) the volume change during phase change process is neglected, i.e., $\rho_f = \rho_l = \rho_s$; (5) the thermal dispersion effects and surface tension are neglected. To take the non-Darcy effect of inertial and viscous forces into consideration, the flow field is described by the generalized non-Darcy model (also called the Brinkman-Forchheimer extended Darcy model) [44-46]. The volume-averaged mass and momentum conservation equations of the generalized non-Darcy model can be written as

$$\nabla \cdot \mathbf{u} = 0 , \tag{1}$$

$$\frac{\partial \mathbf{u}}{\partial t} + (\mathbf{u} \cdot \nabla)\left(\frac{\mathbf{u}}{\phi}\right) = -\frac{1}{\rho_f}\nabla(\phi p) + \nu_e \nabla^2 \mathbf{u} + \mathbf{F} , \tag{2}$$

where $\rho_f$ is the density of the PCM, $\mathbf{u}$ and $p$ are the volume-averaged velocity and pressure,

respectively, $\phi$ is the porosity of the metal foam, $v_e$ is the effective kinematic viscosity, and **F** is the total body force induced by the porous matrix (metal foam) and other external force fields, which can be expressed as [45,46]

$$\mathbf{F} = -\frac{\phi v_f}{K}\mathbf{u} - \frac{\phi F_\phi}{\sqrt{K}}|\mathbf{u}|\mathbf{u} + \phi \mathbf{G}, \quad (3)$$

where $K$ is the permeability, $v_f$ is the kinematic viscosity of the PCM ($v_f$ is not necessarily the same as $v_e$), and **G** is the buoyancy force. The inertial coefficient $F_\phi$ (Forchheimer coefficient) and permeability $K$ depend on the geometry of the metal foam. For flow over a packed bed of particles, based on Ergun's experimental investigations [47], $F_\phi$ and $K$ can be expressed as [48]

$$F_\phi = \frac{1.75}{\sqrt{150\phi^3}}, \quad K = \frac{\phi^3 d_p^2}{150(1-\phi)^2}, \quad (4)$$

where $d_p$ is the solid particle diameter (or mean pore diameter). For metal foam with $\phi = 0.8$ considered in the present study, $F_\phi$ is set to be 0.068 [15,49].

The LTNE model is employed to take into account the temperature differences between metal foam and PCM. According to Refs. [15,17,20], the energy equations of the PCM (including liquid and solid phases) and the metal foam can be written as follows

$$\frac{\partial}{\partial t}\left\{\phi\left[f_l \rho_l c_{pl} + (1-f_l)\rho_s c_{ps}\right]T_f\right\} + \nabla \cdot (\rho_l c_{pl} T_f \mathbf{u}) = \nabla \cdot (\phi k_f \nabla T_f) + h_v(T_m - T_f) - \underline{\frac{\partial}{\partial t}(\phi \rho_l L_a f_l)}, \quad (5)$$

$$\frac{\partial}{\partial t}\left[(1-\phi)\rho_m c_{pm} T_m\right] = \nabla \cdot \left[(1-\phi)k_m \nabla T_m\right] + h_v(T_f - T_m), \quad (6)$$

respectively, where $T$ is the temperature, $f_l$ is the fraction of liquid in PCM ($f_l = 0$ represents the solid phase, $f_l = 1$ represents the liquid phase, and $0 < f_l < 1$ represents the interface region or mushy zone), $c_p$ is the specific heat, $k$ is the thermal conductivity, $h_v = h_{mf} a_{mf}$ is the volumetric heat transfer coefficient ($h_{mf}$ is the interfacial heat transfer coefficient between PCM and metal foam, $a_{mf}$ is the specific surface area of the metal foam [15]), and $L_a$ is the latent heat of phase change. The underlined term in Eq. (5) is the nonlinear latent heat source term accounting for the phase change.

Based on the Boussinesq approximation, the buoyancy force $\mathbf{G}$ in Eq. (3) is given by

$$\mathbf{G} = -\mathbf{g}\beta(T_f - T_0)f_l, \tag{7}$$

where $\mathbf{g}$ is the gravitational acceleration, $\beta$ is the thermal expansion coefficient, and $T_0$ is the reference temperature. The effective thermal conductivities of the PCM and metal foam are defined by

$$k_{e,f} = \phi k_f, \quad k_{e,m} = (1-\phi)k_m, \tag{8}$$

respectively. The thermal conductivity and specific heat of the PCM are given as follows

$$k_f = f_l k_l + (1-f_l)k_s, \quad c_{pf} = f_l c_{pl} + (1-f_l)c_{ps}. \tag{9}$$

Under the local thermal equilibrium (LTE) condition, i.e., $T_f = T_m = T$, the energy equations (5) and (6) can be replaced by the following single-temperature equation [50]

$$\frac{\partial}{\partial t}(\overline{\rho c_p}T) + \nabla \cdot (\rho_l c_{pl} T\mathbf{u}) = \nabla \cdot (k_e \nabla T) - \frac{\partial}{\partial t}(\phi \rho_l L_a f_l), \tag{10}$$

where $\overline{\rho c_p} = \phi[f_l \rho_l c_{pl} + (1-f_l)\rho_s c_{ps}] + (1-\phi)\rho_m c_{pm}$, and $k_e = \phi k_f + (1-\phi)k_m$.

## III. ENTHALPY-BASED MRT-LB METHOD

The LB method has been proved to be a promising method for simulating solid-liquid phase change due to its distinctive advantages (see Ref. [43] for details). In the LB community, the first attempt to use LB method to study solid-liquid phase change was made by De Fabritiis et al. [51] in 1998. Since then, many LB models for solid-liquid phase change have been developed from different points of view [21-23,52-70]. The existing LB models for solid-liquid phase change mostly fall into one of the following categories: the phase-field method [52-59] and the enthalpy-based method [21-23, 60-68]. Additionally, a couple of LB models were recently developed based on some interfacial tracking methods [69,70]. Owing to its simplicity and effectiveness, the enthalpy-based method plays an increasingly important role in simulating solid-liquid phase change problems.

In what follows, an MRT-LB method in conjunction with the enthalpy method will be presented for solid-liquid phase change heat transfer in metal foams under LTNE condition. The method is constructed in the framework of the triple-distribution-function (TDF) approach: the flow field, the temperature fields of PCM and metal foam are solved separately by three different MRT-LB models. For two-dimensional (2D) problems considered in the present study, the two-dimensional nine-velocity (D2Q9) lattice is employed. The nine discrete velocities $\{\mathbf{e}_i\}$ of the D2Q9 lattice are given by [27]

$$\mathbf{e}_i = \begin{cases} (0,0), & i=0, \\ \left(\cos\left[(i-1)\pi/2\right], \sin\left[(i-1)\pi/2\right]\right)c, & i=1\sim 4, \\ \left(\cos\left[(2i-9)\pi/4\right], \sin\left[(2i-9)\pi/4\right]\right)\sqrt{2}c, & i=5\sim 8, \end{cases} \tag{11}$$

where $c = \delta_x/\delta_t$ is the lattice speed with $\delta_t$ and $\delta_x$ being the discrete time step and lattice spacing, respectively.

### A. MRT-LB model for flow field

The MRT method [28,71] is an important extension of the relaxation LB method developed by Higuera et al. [25]. In MRT method, the collision process of the LB equation is executed in moment space, while the streaming process of the LB equation is carried out in velocity space. By using the MRT collision model, the relaxation times of the hydrodynamic and non-hydrodynamic moments can be separated. According to Ref. [72,73], the MRT-LB equation with an explicit treatment of the forcing term can be written as

$$f_i^+\left(\mathbf{x}+\mathbf{e}_i\delta_t, t+\delta_t\right) = f_i(\mathbf{x},t) - \tilde{\Lambda}_{ij}\left(f_j - f_j^{eq}\right)\Big|_{(\mathbf{x},t)} + \delta_t\left(\tilde{S}_i - 0.5\tilde{\Lambda}_{ij}\tilde{S}_j\right), \tag{12}$$

where $f_i(\mathbf{x},t)$ is the (volume-averaged) density distribution function, $f_i^{eq}(\mathbf{x},t)$ is the equilibrium distribution function, $\tilde{S}_i$ is the forcing term, and $\tilde{\Lambda} = \mathbf{M}^{-1}\Lambda\mathbf{M}$ is the collision matrix in velocity space. Here, $\mathbf{M}$ is the transformation matrix, and $\Lambda$ is the relaxation matrix. For the D2Q9 model, the transformation matrix $\mathbf{M}$ is given by [71]

$$\mathbf{M} = \begin{bmatrix} 1 & 1 & 1 & 1 & 1 & 1 & 1 & 1 & 1 \\ -4 & -1 & -1 & -1 & -1 & 2 & 2 & 2 & 2 \\ 4 & -2 & -2 & -2 & -2 & 1 & 1 & 1 & 1 \\ 0 & 1 & 0 & -1 & 0 & 1 & -1 & -1 & 1 \\ 0 & -2 & 0 & 2 & 0 & 1 & -1 & -1 & 1 \\ 0 & 0 & 1 & 0 & -1 & 1 & 1 & -1 & -1 \\ 0 & 0 & -2 & 0 & 2 & 1 & 1 & -1 & -1 \\ 0 & 1 & -1 & 1 & -1 & 0 & 0 & 0 & 0 \\ 0 & 0 & 0 & 0 & 0 & 1 & -1 & 1 & -1 \end{bmatrix}. \tag{13}$$

Through the transformation matrix $\mathbf{M}$, the collision process of the MRT-LB equation (12) can be executed in moment space $\mathbb{M} = \mathbb{R}^9$, i.e.,

$$\mathbf{m}^*(\mathbf{x}, t) = \mathbf{m}(\mathbf{x}, t) - \mathbf{\Lambda}(\mathbf{m} - \mathbf{m}^{eq})\big|_{(\mathbf{x}, t)} + \delta_t \left(\mathbf{I} - \frac{\mathbf{\Lambda}}{2}\right)\mathbf{S}, \tag{14}$$

where the bold-face symbols $\mathbf{m}$, $\mathbf{m}^{eq}$, and $\mathbf{S}$ denote 9-dimensional column vectors of moments, e.g., $\mathbf{m} = (m_0, m_1, \ldots, m_8)^\mathrm{T}$. The streaming process is still carried out in velocity space $\mathbb{V} = \mathbb{R}^9$

$$f_i^+(\mathbf{x} + \mathbf{e}_i \delta_t, t + \delta_t) = f_i^*(\mathbf{x}, t), \tag{15}$$

where $\mathbf{f}^* = \mathbf{M}^{-1}\mathbf{m}^*$. The superscript "+" denotes that the effect of the solid phase has not yet been considered. The diagonal relaxation matrix $\mathbf{\Lambda}$ is given by

$$\mathbf{\Lambda} = \mathrm{diag}(s_\rho, s_e, s_\varepsilon, s_j, s_q, s_j, s_q, s_v, s_v). \tag{16}$$

The transformation matrix $\mathbf{M}$ linearly maps the discrete distribution functions represented by $\mathbf{f} \in \mathbb{V} = \mathbb{R}^9$ to their velocity moments represented by $\mathbf{m} \in \mathbb{M} = \mathbb{R}^9$, as in the following

$$\mathbf{m} = \mathbf{Mf}, \quad \mathbf{f} = \mathbf{M}^{-1}\mathbf{m}. \tag{17}$$

The equilibrium moment $\mathbf{m}^{eq}$ corresponding to $\mathbf{m}$ is defined as [73]

$$\mathbf{m}^{eq} = \rho\left(1, -2 + \frac{3|\mathbf{u}|^2}{\phi c^2}, \alpha_1 + \frac{\alpha_2|\mathbf{u}|^2}{\phi c^2}, \frac{u_x}{c}, -\frac{u_x}{c}, \frac{u_y}{c}, -\frac{u_y}{c}, \frac{u_x^2 - u_y^2}{\phi c^2}, \frac{u_x u_y}{\phi c^2}\right)^\mathrm{T}, \tag{18}$$

where $\rho = \rho_f$, and $\alpha_1$ and $\alpha_2$ are free parameters. The forcing term in moment space $\mathbf{S}$ is given by [73]

$$\mathbf{S} = \rho \left( 0, \frac{6\mathbf{u}\cdot\mathbf{F}}{\phi c^2}, -\frac{6\mathbf{u}\cdot\mathbf{F}}{\phi c^2}, \frac{F_x}{c}, -\frac{F_x}{c}, \frac{F_y}{c}, -\frac{F_y}{c}, \frac{2(u_x F_x - u_y F_y)}{\phi c^2}, \frac{u_x F_y + u_y F_x}{\phi c^2} \right)^{\mathrm{T}}, \tag{19}$$

where $F_x$ and $F_y$ are x- and y-components of the total body force $\mathbf{F}$, respectively.

As mentioned in Section I, it is not appropriate to use the bounce-back scheme (liquid fraction $f_l = 0.5$ is defined as the phase interface, the collision process (14) is performed for $f_l > 0.5$) [62] to impose the no-slip velocity condition in the interface region. To accurately realize the no-slip velocity condition in the interface and solid phase regions, the volumetric LB scheme [67] is employed in the present study. By using the volumetric LB scheme, the flow field is modeled over the entire domain (including liquid and solid phase regions). Considering the effect of the solid phase, the density distribution function $f_i$ is redefined as

$$f_i = f_l f_i^+ + (1 - f_l) f_i^{eq}(\rho, \mathbf{u}_s), \tag{20}$$

where $f_i^+$ is given by Eq. (15), and $\mathbf{u}_s = 0$ is the velocity of the solid phase. The above equation is based on a kinetic assumption that the solid phase density distribution function is at equilibrium state. Accordingly, the macroscopic density $\rho$ and velocity $\mathbf{u}$ are defined as

$$\rho = \sum_{i=0}^{8} f_i^+, \tag{21}$$

$$\rho \mathbf{u} = \sum_{i=0}^{8} \mathbf{e}_i f_i + \frac{\delta_t}{2} \rho \mathbf{F}. \tag{22}$$

The macroscopic pressure $p$ is given by $p = \rho c_s^2 / \phi$. Eq. (22) is a nonlinear equation for the velocity $\mathbf{u}$ because $\mathbf{F}$ also contains the velocity. According to Ref. [74], the macroscopic velocity $\mathbf{u}$ can be calculated explicitly by

$$\mathbf{u} = \frac{\mathbf{v}}{l_0 + \sqrt{l_0^2 + l_1 |\mathbf{v}|}}, \tag{23}$$

where

$$\rho \mathbf{v} = \sum_{i=0}^{8} \mathbf{e}_i f_i + \frac{\delta_t}{2} \phi \rho \mathbf{G}, \tag{24}$$

$$l_0 = \frac{1}{2}\left(1 + \phi \frac{\delta_t}{2} \frac{v_f}{K}\right), \quad l_1 = \phi \frac{\delta_t}{2} \frac{F_\phi}{\sqrt{K}}. \tag{25}$$

Through the Chapman-Enskog analysis of the MRT-LB equation (12), the mass and momentum conservation equations (1) and (2) can be recovered in the incompressible limit. The effective kinematic viscosity $v_e$ and the bulk viscosity $v_B$ are given by

$$v_e = c_s^2 \left(\frac{1}{s_v} - \frac{1}{2}\right)\delta_t, \quad v_B = c_s^2 \left(\frac{1}{s_e} - \frac{1}{2}\right)\delta_t, \tag{26}$$

respectively, where $s_{7,8} = s_v = 1/\tau_v$ ($\tau_v$ is the relaxation time), $c_s = c/\sqrt{3}$ is the sound speed of the D2Q9 model.

The equilibrium distribution function $f_i^{eq}$ in velocity space is given by ($\alpha_1 = 1$, $\alpha_2 = -3$) [74]

$$f_i^{eq} = w_i \rho \left[1 + \frac{\mathbf{e}_i \cdot \mathbf{u}}{c_s^2} + \frac{(\mathbf{e}_i \cdot \mathbf{u})^2}{2\phi c_s^4} - \frac{|\mathbf{u}|^2}{2\phi c_s^2}\right], \tag{27}$$

where $w_0 = 4/9$, $w_{1\sim4} = 1/9$, and $w_{5\sim8} = 1/36$.

### B. MRT-LB models for temperature fields

For solid-liquid phase change heat transfer in metal foams under LTNE condition, the temperature fields are solved separately by two different MRT-LB models: an enthalpy-based MRT-LB model is proposed to solve the PCM temperature field, while an internal-energy-based MRT-LB model is proposed to solve the metal foam temperature field. In this subsection, the MRT-LB models for temperature fields will be presented. In addition, some remarks about the MRT-LB models will also be presented.

#### *1. Enthalpy-based MRT-LB model for PCM temperature field*

By combining the nonlinear latent heat source term $\partial_t (\phi \rho_l L_a f_l)$ into the transient term in Eq. (5), the following enthalpy-based energy equation of the PCM can be obtained

$$\frac{\partial H_f}{\partial t} + \nabla \cdot \left( \frac{c_{pl} T_f \mathbf{u}}{\phi} \right) = \nabla \cdot \left( \frac{k_f}{\rho_l} \nabla T_f \right) + \frac{h_v (T_m - T_f)}{\phi \rho_l}, \tag{28}$$

where $H_f$ is the enthalpy of the PCM. The enthalpy $H_f$ can be divided into two parts: the sensible enthalpy $c_{pf} T_f$ and the latent enthalpy $f_l L_a$, i.e.,

$$H_f = c_{pf} T_f + f_l L_a = f_l H_l + (1 - f_l) H_s, \tag{29}$$

$$H_l = c_{pl} T_f + L_a, \quad H_s = c_{ps} T_f, \tag{30}$$

where $H_l$ is the enthalpy of the liquid PCM, and $H_s$ is the enthalpy of the solid PCM.

For the PCM temperature field governed by Eq. (28), the following MRT-LB equation of the enthalpy distribution function $g_i(\mathbf{x}, t)$ is introduced

$$\mathbf{g}(\mathbf{x} + \mathbf{e}\delta_t, t + \delta_t) = \mathbf{g}(\mathbf{x}, t) - \mathbf{M}^{-1} \mathbf{\Theta} \left( \mathbf{n}_g - \mathbf{n}_g^{eq} \right)\Big|_{(\mathbf{x}, t)} + \delta_t \mathbf{M}^{-1} \mathbf{S}_{\text{PCM}}, \tag{31}$$

where $\mathbf{M}$ is the transformation matrix [see Eq. (13)], and $\mathbf{\Theta} = \text{diag}(\zeta_0, \zeta_e, \zeta_\varepsilon, \zeta_\alpha, \zeta_q, \zeta_\alpha, \zeta_q, \zeta_v, \zeta_v)$ is the relaxation matrix. The collision process of the above MRT-LB equation is executed in moment space, i.e.,

$$\mathbf{n}_g^*(\mathbf{x}, t) = \mathbf{n}_g(\mathbf{x}, t) - \mathbf{\Theta} \left( \mathbf{n}_g - \mathbf{n}_g^{eq} \right)\Big|_{(\mathbf{x}, t)} + \delta_t \mathbf{S}_{\text{PCM}}, \tag{32}$$

where $\mathbf{n}_g = \mathbf{M} \mathbf{g}$ is the moment, and $\mathbf{n}_g^{eq} = \mathbf{M} \mathbf{g}^{eq}$ is the corresponding equilibrium moment. Here, $g_i^{eq}$ is the equilibrium enthalpy distribution function in velocity space. The streaming process is carried out in velocity space

$$g_i(\mathbf{x} + \mathbf{e}_i \delta_t, t + \delta_t) = g_i^*(\mathbf{x}, t), \tag{33}$$

where $\mathbf{g}^* = \mathbf{M}^{-1} \mathbf{n}_g^*$.

The equilibrium moment $\mathbf{n}_g^{eq}$ can be chosen as

$$\mathbf{n}_g^{eq} = \left( H_f, -4H_f + 2c_{f,\text{ref}} T_f, 4H_f - 3c_{f,\text{ref}} T_f, \frac{c_{pl} T_f u_x}{\phi c}, -\frac{c_{pl} T_f u_x}{\phi c}, \frac{c_{pl} T_f u_y}{\phi c}, -\frac{c_{pl} T_f u_y}{\phi c}, 0, 0 \right)^{\text{T}}, \tag{34}$$

where $c_{f,\text{ref}}$ is a reference specific heat. As did in Ref. [66], the reference specific heat is introduced into the equilibrium moment to make the specific heat and thermal conductivity of the PCM decoupled.

To recover the enthalpy-based energy equation (28), the source term in moment space $\mathbf{S}_{PCM}$ is chosen as

$$\mathbf{S}_{PCM} = S_{PCM}(1,-2,1,0,0,0,0,0,0)^{\mathsf{T}}, \tag{35}$$

where $S_{PCM}$ is given by

$$S_{PCM} = Sr_f + \frac{1}{2}\delta_t \partial_t Sr_f, \quad Sr_f = \frac{h_v(T_m - T_f)}{\phi \rho_l}. \tag{36}$$

The enthalpy-based energy equation (28) is actually a nonlinear convection-diffusion equation with a source term. Therefore, a time derivative term $\frac{1}{2}\delta_t \partial_t Sr_f$ is incorporated into the source term $\mathbf{S}_{PCM}$ as suggested in the literature [75]. Without this derivative term, there must exist an unwanted term $\frac{\delta_t}{2}\epsilon \partial_{t_1} Sr_f$ in the macroscopic equation recovered from the MRT-LB equation (31). The details will be described later through the Chapman-Enskog analysis [76] in Appendix A.

The enthalpy $H_f$ is computed by

$$H_f = n_{g0} = \sum_{i=0}^{8} g_i. \tag{37}$$

The relationship between the enthalpy $H_f$ and temperature $T_f$ is given by

$$T_f = \begin{cases} H_f/c_{ps}, & H_f \leq H_{fs}, \\ T_{fs} + \dfrac{H_f - H_{fs}}{H_{fl} - H_{fs}}(T_{fl} - T_{fs}), & H_{fs} < H_f < H_{fl}, \\ T_{fl} + (H_f - H_{fl})/c_{pl}, & H_f \geq H_{fl}, \end{cases} \tag{38}$$

where $H_{fs} = c_{ps}T_{fs}$ is the enthalpy at the solidus temperature $T_{fs}$, and $H_{fl} = c_{pl}T_{fl} + L_a$ is the enthalpy at the liquidus temperature $T_{fl}$. The liquid fraction $f_l$ can be determined by

$$f_l = \begin{cases} 0, & H_f \leq H_{fs}, \\ \dfrac{H_f - H_{fs}}{H_{fl} - H_{fs}}, & H_{fs} < H_f < H_{fl}, \\ 1, & H_f \geq H_{fl}. \end{cases} \tag{39}$$

The equilibrium enthalpy distribution function $g_i^{eq}$ in velocity space is given by

$$g_i^{eq} = \begin{cases} H_f - \dfrac{5}{9} c_{f,\text{ref}} T_f, & i = 0, \\ w_i c_{pl} T_f \left( \dfrac{c_{f,\text{ref}}}{c_{pl}} + \dfrac{\mathbf{e}_i \cdot \mathbf{u}}{\phi c_s^2} \right), & i = 1 \sim 8. \end{cases} \quad (40)$$

## *2. Internal-energy-based MRT-LB model for metal foam temperature field*

The energy equation (6) of the metal foam can be rewritten as

$$\frac{\partial (c_{pm} T_m)}{\partial t} = \nabla \cdot \left( \frac{k_m}{\rho_m} \nabla T_m \right) + \frac{h_v (T_f - T_m)}{(1-\phi) \rho_m}. \quad (41)$$

For the metal foam temperature field governed by the above equation, the MRT-LB equation of the internal-energy distribution function $h_i(\mathbf{x}, t)$ is given by

$$\mathbf{h}(\mathbf{x} + \mathbf{e}\delta_t, t + \delta_t) = \mathbf{h}(\mathbf{x}, t) - \mathbf{M}^{-1} \mathbf{Q} \left( \mathbf{n}_h - \mathbf{n}_h^{eq} \right) \Big|_{(\mathbf{x}, t)} + \delta_t \mathbf{M}^{-1} \mathbf{S}_{\text{metal}}, \quad (42)$$

where $\mathbf{Q} = \text{diag}(\eta_0, \eta_e, \eta_\varepsilon, \eta_\alpha, \eta_q, \eta_\alpha, \eta_q, \eta_v, \eta_v)$ is the relaxation matrix. The collision process of the above MRT-LB equation is executed in moment space, i.e.,

$$\mathbf{n}_h^*(\mathbf{x}, t) = \mathbf{n}_h(\mathbf{x}, t) - \mathbf{Q} \left( \mathbf{n}_h - \mathbf{n}_h^{eq} \right) \Big|_{(\mathbf{x}, t)} + \delta_t \mathbf{S}_{\text{metal}}, \quad (43)$$

where $\mathbf{n}_h = \mathbf{M}\mathbf{h}$ is the moment, and $\mathbf{n}_h^{eq} = \mathbf{M}\mathbf{h}^{eq}$ is the corresponding equilibrium moment. Here, $h_i^{eq}$ is the equilibrium internal-energy distribution function in velocity space. The streaming process is carried out in velocity space

$$h_i(\mathbf{x} + \mathbf{e}_i \delta_t, t + \delta_t) = h_i^*(\mathbf{x}, t), \quad (44)$$

where $\mathbf{h}^* = \mathbf{M}^{-1} \mathbf{n}_h^*$.

The equilibrium moment $\mathbf{n}_h^{eq}$ is defined as

$$\mathbf{n}_h^{eq} = \left( c_{pm} T_m, -4 c_{pm} T_m + 2 c_{m,\text{ref}} T_m, 4 c_{pm} T_m - 3 c_{m,\text{ref}} T_m, 0, 0, 0, 0, 0, 0 \right)^\mathrm{T}, \quad (45)$$

where $c_{m,\text{ref}}$ is a reference specific heat. The source term in moment space $\mathbf{S}_{\text{metal}}$ is chosen as

$$\mathbf{S}_{\text{metal}} = S_{\text{metal}} (1, -2, 1, 0, 0, 0, 0, 0, 0)^\mathrm{T}, \quad (46)$$

where $S_{\text{metal}}$ is given by

$$S_{\text{metal}} = Sr_m + \frac{1}{2}\delta_t \partial_t Sr_m, \quad Sr_m = \frac{h_v(T_f - T_m)}{(1-\phi)\rho_m}. \tag{47}$$

The temperature $T_m$ is defined by

$$T_m = \frac{1}{c_{pm}}\sum_{i=0}^{8} h_i. \tag{48}$$

The equilibrium internal-energy distribution function $h_i^{eq}$ in velocity space is given by

$$h_i^{eq} = \begin{cases} c_{pm}T_m - \dfrac{5}{9}c_{m,\text{ref}}T_m, & i=0, \\ w_i c_{m,\text{ref}}T_m, & i=1\sim 8. \end{cases} \tag{49}$$

Through the Chapman-Enskog analysis [76] of the MRT-LB equation (31), the following macroscopic energy equation can be obtained (see Appendix A for details)

$$\frac{\partial H_f}{\partial t} + \nabla \cdot \left(\frac{c_{pl}T_f \mathbf{u}}{\phi}\right) = \nabla \cdot \left[\alpha_{f,\text{ref}}c_{f,\text{ref}}\nabla T_f + \delta_t\left(\zeta_\alpha^{-1} - 0.5\right)\epsilon \partial_{t_1}\left(\frac{c_{pl}T_f \mathbf{u}}{\phi}\right)\right] + Sr_f, \tag{50}$$

where $\zeta_{3,5} = \zeta_\alpha = 1/\tau_g$ ($\tau_g$ is the relaxation time), and $\alpha_{f,\text{ref}}$ is the reference thermal diffusivity [see Eq. (A19)]. As compared with the enthalpy-based energy equation (28), Eq. (50) contains an additional term $\nabla \cdot \left[\delta_t\left(\zeta_\alpha^{-1} - 0.5\right)\epsilon \partial_{t_1}\left(c_{pl}T_f \mathbf{u}/\phi\right)\right]$. For incompressible thermal flows, the additional term can be neglected in most cases, then the enthalpy-based energy equation of the PCM can be asymptotically recovered from the MRT-LB equation (31). Similarly, the energy equation of the metal foam can be asymptotically recovered from the MRT-LB equation (42) as

$$\frac{\partial(c_{pm}T_m)}{\partial t} = \nabla \cdot \left(\alpha_{m,\text{ref}}c_{m,\text{ref}}\nabla T_m\right) + Sr_m, \tag{51}$$

where $\alpha_{m,\text{ref}} = k_m/(\rho_m c_{m,\text{ref}}) = c_s^2\left(\eta_\alpha^{-1} - 0.5\right)\delta_t$ is the reference thermal diffusivity with $\eta_{3,5} = \eta_\alpha = 1/\tau_h$ ($\tau_h$ is the relaxation time).

In this subsection, the MRT-LB models for the temperature fields have been developed based on the LTNE model. In what follows, some remarks are presented on the proposed models.

*Remark I.* The reference specific heats $c_{f,\text{ref}}$ and $c_{m,\text{ref}}$ keep unvaried over the entire domain,

which makes the thermal conductivity and specific heat of the PCM (or metal foam) decoupled. As a result, the differences in specific heat and thermal conductivity can be naturally handled [see Eqs. (A17) and (A18)]. According to Eqs. (40) and (49), $c_{f,\text{ref}}$ and $c_{m,\text{ref}}$ should satisfy $c_{f,\text{ref}} < \frac{9}{5} c_{pf}$ and $c_{m,\text{ref}} < \frac{9}{5} c_{pm}$, respectively.

*Remark II*. For solid-liquid phase change without convective effect, i.e., the velocity $\mathbf{u}$ is zero, the additional term in Eq. (50) disappears. For solid-liquid phase change coupled with natural convection, the additional term has no effect on numerical simulations in most cases, thus it has been neglected in the present study. Theoretically, to remove the additional term, the approaches in Refs. [67, 75] can be employed.

*Remark III*. The energy equations (28) and (41) are nonlinear convection-diffusion equations with source terms. Therefore, time derivative terms [see Eqs. (36) and (47)] are incorporated into the MRT-LB equations for the temperature fields. Without the derivative terms, there must exist unwanted terms in the macroscopic equations recovered from the MRT-LB equations (31) and (42), as can be seen from Eqs. (A6) and (A8). In simulations, the explicit difference scheme can be used to compute the time derivative terms (e.g., $\partial_t Sr_f = \left[ Sr_f(\mathbf{x},t) - Sr_f(\mathbf{x},t-\delta_t) \right]/\delta_t$), which does not affect the inherent merits of the LB method. Unlike the iteration method in previous studies [21-23], the MRT-LB equation (31) is completely local and is easy to implement in the same way as the standard MRT-LB equation.

*Remark IV*. The two-dimensional five-velocity (D2Q5) lattice can also be employed. The MRT-LB models for the temperature fields based on D2Q5 lattice are presented in Appendix B.

### C. Boundary conditions and relaxation rates

In this subsection, the boundary conditions and relaxation rates are briefly introduced. For velocity and thermal boundary conditions, the non-equilibrium extrapolation scheme [77] is employed. It should be noted that the no-slip velocity boundary condition on the walls is treated based on $f_i^+$ rather than $f_i$, i.e., they are treated before the consideration of the effect of the solid phase. For a boundary node $\mathbf{x}_b$ where $\mathbf{u}(\mathbf{x}_b, t)$ is known, but $\rho(\mathbf{x}_b, t)$ is unknown, the discrete density distribution function $f_i^+(\mathbf{x}_b, t)$ at the boundary node $\mathbf{x}_b$ is given by

$$f_i^+(\mathbf{x}_b, t) = \hat{f}_i^{eq}(\mathbf{x}_b, t) + \left[ f_i^+(\mathbf{x}_f, t) - f_i^{eq}(\mathbf{x}_f, t) \right], \tag{52}$$

where $\hat{f}_i^{eq}(\mathbf{x}_b, t) = f_i^{eq}(\rho(\mathbf{x}_f), \mathbf{u}(\mathbf{x}_b), t)$, and $\mathbf{x}_f$ is the nearest neighbor fluid node of $\mathbf{x}_b$ along the link $\mathbf{e}_i$, i.e., $\mathbf{x}_f = \mathbf{x}_b + \mathbf{e}_i \delta_t$.

In the MRT-LB model for flow field, the relaxation rate (related to the effective kinematic viscosity) $s_v$ is determined by $s_v^{-1} = 0.5 + v_e / (c_s^2 \delta_t)$, the free relaxation rates are set as $s_\rho = s_j = 1$, $s_e = s_\varepsilon = 1.1$, $s_q = 1.2$. In the MRT-LB models for temperature fields, the relaxation rates (related to the reference thermal diffusivities) $\zeta_\alpha$ and $\eta_\alpha$ are determined by $\zeta_\alpha^{-1} = 0.5 + \alpha_{f,\text{ref}} / (c_s^2 \delta_t)$ and $\eta_\alpha^{-1} = 0.5 + \alpha_{m,\text{ref}} / (c_s^2 \delta_t)$, respectively, the free relaxation rates are set as $\zeta_e = \zeta_\varepsilon = 2 - \zeta_\alpha$, $\zeta_q = \zeta_\alpha$, $\zeta_v = 1.2$, $\eta_0 = 1$, $\eta_e = \eta_\varepsilon = 1.1$, $\eta_q = \eta_\alpha$, $\eta_v = 1.2$. We would also like to point out that the relaxation rate $\zeta_e$ plays an important role in simulating solid-liquid phase change heat transfer in metal foams under LTNE condition. To reduce numerical diffusion across phase interface, we set $\zeta_e = 2 - \zeta_\alpha$, i.e.,

$$\left( \frac{1}{\zeta_e} - \frac{1}{2} \right)\left( \frac{1}{\zeta_\alpha} - \frac{1}{2} \right) = \frac{1}{4}. \tag{53}$$

As reported in Ref. [66], by using the above relationship, the numerical diffusion across the phase interface can be significantly reduced in simulating solid-liquid phase change problems without porous media. Although solid-liquid phase change heat transfer in metal foams is much more complicated, the relationship given by Eq. (53) is employed in the present study.

## IV. NUMERICAL TESTS

In this section, numerical simulations of two solid-liquid phase change heat transfer problems in metal foams under LTNE condition are carried out to validate the accuracy and effectiveness of the enthalpy-based MRT-LB method. The problems are characterized by the following dimensionless parameters: Rayleigh number $Ra$, Prandtl number $Pr$, Darcy number $Da$, viscosity ratio $J$, metal foam-to-PCM thermal conductivity ratio $\lambda$, metal foam-to-PCM thermal diffusivity ratio $\Gamma$, metal foam-to-PCM heat capacity ratio $\hat{\sigma}$, volumetric heat transfer coefficient $H_v$ (based on pore diameter $d_p$), Fourier number $Fo$ (dimensionless time), and Stefan number $St$, which are defined as follows

$$Ra = \frac{g\beta\Delta T L^3}{v_f \alpha_f}, \quad Pr = \frac{v_f}{\alpha_f}, \quad Da = \frac{K}{L^2}, \quad J = \frac{v_e}{v_f}, \quad \lambda = \frac{k_m}{k_f},$$

$$\Gamma = \frac{\alpha_m}{\alpha_f}, \quad \hat{\sigma} = \frac{(\rho c_p)_m}{(\rho c_p)_f}, \quad H_v = \frac{h_v d_p^2}{k_f}, \quad Fo = \frac{t\alpha_f}{L^2}, \quad St = \frac{c_{pl}\Delta T}{L_a}, \tag{54}$$

where $L$ is the characteristic length, $\Delta T$ is the characteristic temperature, $\alpha_f = k_f/(\rho c_p)_f$ and $\alpha_m = k_m/(\rho c_p)_m$ are thermal diffusivities of the PCM and metal foam, respectively.

Some required parameters are set as follows: $\phi = 0.8$, $F_\phi = 0.068$, $c_{pl} = c_{ps} = 1$, $c_{pf} = c_{pm} = 1$, $\hat{\sigma} = 1$, $\rho_0 = 1$ (reference density of the PCM), $\delta_x = \delta_t = 1$ ($c = 1$). Note that it is no need to restrict $c = 1$ in simulations. In order to make comparisons with previous numerical results, following Ref. [15], the volumetric heat transfer coefficient $H_v$ is held constant at $5.9$ and the pore size $d_p/L$ is set to be $0.0135$.

### A. Solidification by conduction

In this subsection, to validate the MRT-LB models for the temperature fields, conduction-induced solidification in a semi-infinite domain is considered. The schematic diagram of this problem is shown in Fig. 1. The computational domain is filled with metal-foam-based PCM. This problem is symmetrical about $y = x$. Initially, the PCM is in liquid state at temperature $T_i$ ($T_i > T_{melt}$, here $T_{melt}$ is the melting temperature). At time $t = 0$, the left and bottom walls are lowered to a fixed temperature $T_c$ ($T_c < T_{melt}$), and consequently, solidification begins along the left and bottom surfaces and proceeds into the PCM.

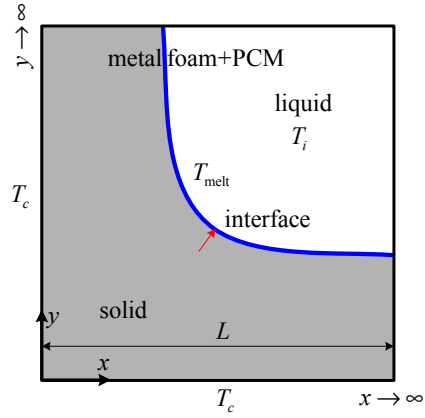

FIG. 1. Schematic diagram of 2D conduction-induced solidification in a semi-infinite domain.

In simulations, the parameters are chosen as follows: $\lambda = 10$, $St = 4$, $T_c = -1$, $T_{melt} = 0$, $T_i = 0.3$, $\Delta T = T_{melt} - T_c = 1$, $k_f = 0.01$, $c_{f,ref} = 0.5c_{pf}$, and $c_{m,ref} = c_{pm}$. The solidus temperature $T_{fs} = -0.05$, and the liquidus temperature $T_{fl} = 0.05$. A grid size of $N_x \times N_y = 200 \times 200$ is employed (the characteristic length $L = N_x/2$), and the velocity field is set to be zero ($\mathbf{u} = 0$) at each lattice node. In Fig. 2, the phase field at $Fo = 0.02$ is shown. The phase change occurs over a range of temperatures, and the phase interface is usually referred as the interface region or mushy zone. The liquid fraction distribution, isotherms of the PCM and metal foam at $Fo = 0.02$ are presented in Fig. 3. As shown in the figure, the gaps of the isotherms in solid phase region are less than those in liquid phase region because of the release of latent heat on the phase interface. For comparison, the results obtained by the finite-difference method (FDM) are also presented in Fig. 3. Obviously, the present results are in good agreement with the FDM results.

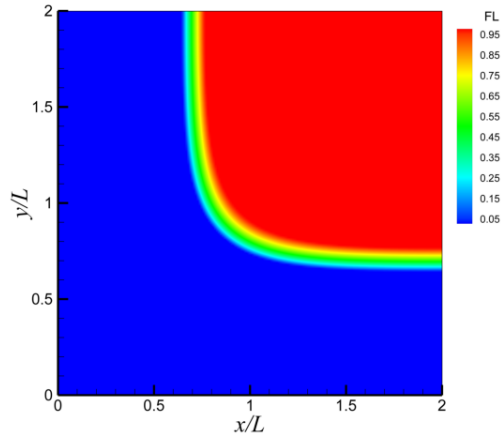

FIG. 2. Phase field of conduction-induced solidification in a semi-infinite domain at Fo = 0.02. The the characteristic length is chosen as $L = N_x/2$.

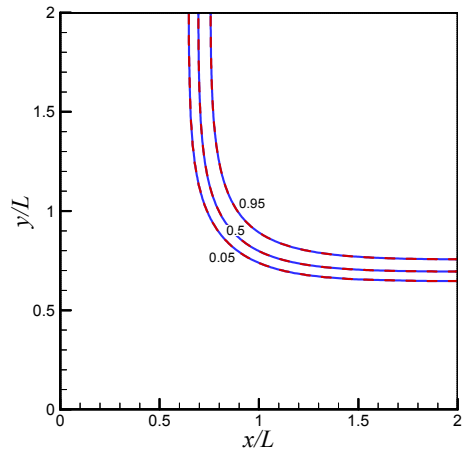

(a) liquid fraction distribution

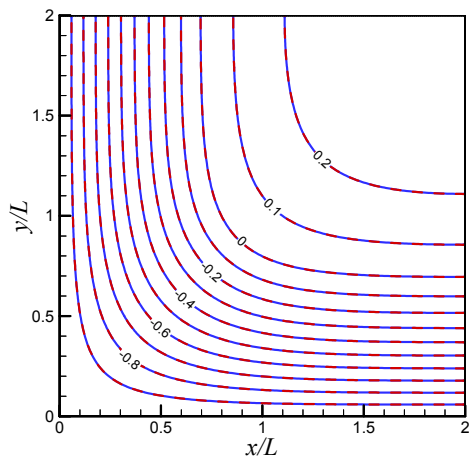

(b) isotherms of the PCM

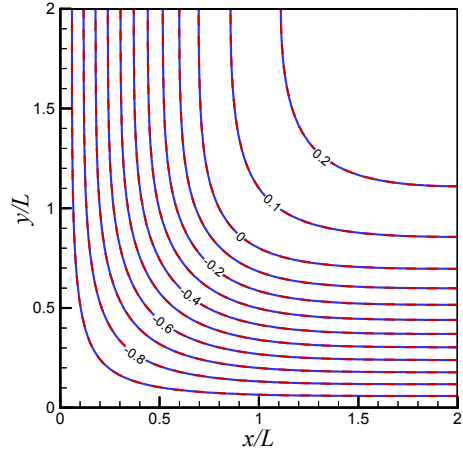

(c) isotherms of the metal foam

FIG. 3. The liquid fraction distribution (a), isotherms of the PCM (b), and metal foam (c) at Fo = 0.02. The blue solid and red dashed lines represent the present and the FDM results, respectively.

### B. Melting coupled with natural convection

In this subsection, numerical simulations of melting coupled with natural convection in a square cavity filled with metal-foam-based PCM are carried out to validate the present method. The schematic diagram of this problem is shown in Fig. 4. The distance between the walls is $L$. The horizontal walls are adiabatic, while the left and right walls are kept at constant temperatures $T_h$ and $T_c$ ($T_h > T_c$), respectively. Initially, the PCM is in solid state at temperature $T_i$ ($T_i < T_{melt}$). At time $t = 0$, the temperature of the left wall is raised to $T_h$ ($T_h > T_{melt}$), and consequently, melting begins along the left wall and proceeds into the PCM inside the cavity.

In simulations, the parameters are set as follows: $Da = 10^{-4}$, $Pr = 50$, $F_\phi = 0.068$, $\lambda = 10^3$, $St = 1$, $J = 1$, $T_h = 1$, $T_{melt} = 0.3$, $T_c = T_i = T_0 = 0$, $\Delta T = T_h - T_c = 1$, $k_f = 0.0005$, $c_{f,ref} = 0.2c_{pl}$, $c_{m,ref} = c_{pm}$. The solidus temperature $T_{fs} = 0.299$, and the liquidus temperature $T_{fl} = 0.301$. For $Ra = 10^6$, a grid size of $N_x \times N_y = 150 \times 150$ is employed, and for $Ra = 10^8$, a grid size of $N_x \times N_y = 300 \times 300$ is employed. In Fig. 5, the locations of the phase interface at different Fourier numbers are presented. For comparison purpose, the locations of the phase interface are determined by

$f_l = 0.5$. It can be seen from the figure that the present results agree well with the FVM solutions [15].

From Fig. 5(a) it can be seen that at $Ra = 10^6$, the heat transfer process is dominated by conduction because the metal foam-to-PCM thermal conductivity ratio is very large ($\lambda = 10^3$), and the shape of the phase interface is almost planar during the melting process. As $Ra$ increases to $10^8$, the effect of natural convection on the shape of the phase interface becomes stronger. As shown in Fig. 5(b), due to the convective effect, the phase interface moves faster near the top wall.

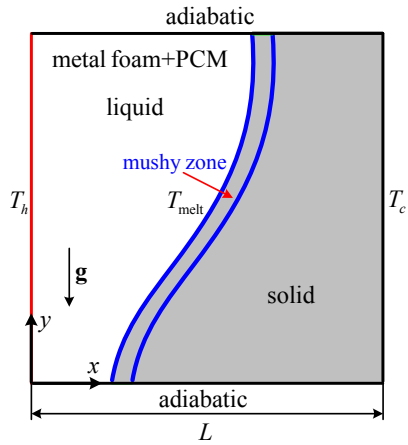

FIG. 4. Schematic diagram of melting coupled with natural convection in a square cavity filled with metal-foam-based PCM.

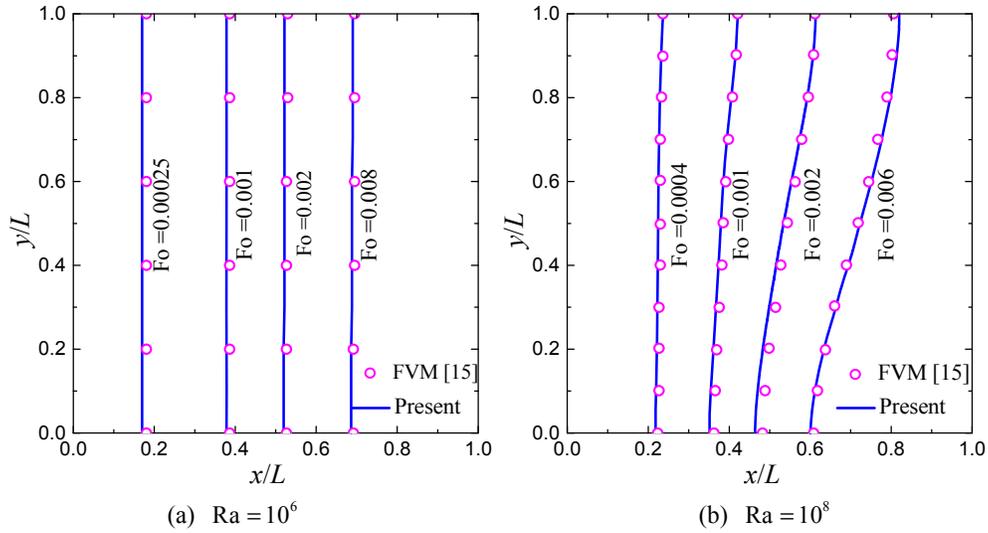

(a) $Ra = 10^6$  (b) $Ra = 10^8$

FIG. 5. Locations of the phase interface for $Ra = 10^6$ (a) and $Ra = 10^8$ (b) at different Fourier numbers. For comparison purpose, the locations of the phase interface are determined by $f_l = 0.5$.

As mentioned in Section I, for solid-liquid phase change in metal-foam-based PCMs, the phase interface is a diffusive interface with a certain thickness rather than a sharp interface, which is usually referred as the interface region or mushy zone. In Fig. 6, the streamlines with the phase interface at different Fourier numbers for $Ra = 10^6$ are shown. From the figure we can see that, during the melting process ($Fo \leq 0.002$), the thickness of the phase interface is around ten lattices, as a result of the interfacial heat transfer between PCM and metal foam. In the quasi-steady regime ($Fo = 0.008$), the movement of the phase interface is slow enough and it only occupies one or two lattices. The streamlines with the phase interface at different Fourier numbers for $Ra = 10^8$ are shown in Fig. 7. The overall behavior is similar to that with $Ra = 10^6$, albeit with stronger convective effect.

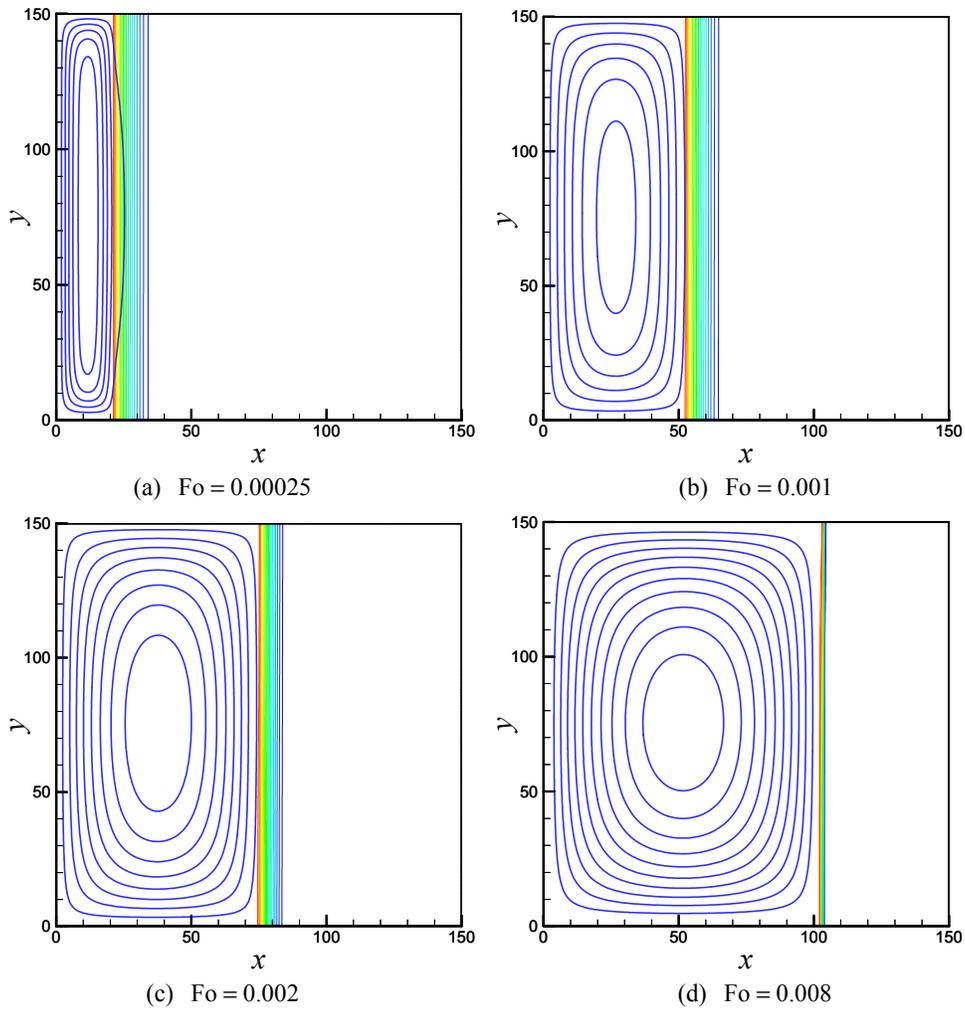

(a) Fo = 0.00025　　　　　　　　　　(b) Fo = 0.001

(c) Fo = 0.002　　　　　　　　　　(d) Fo = 0.008

FIG. 6. Streamlines with the phase interface at different Fourier numbers for $Ra = 10^6$. (a) Fo = 0.00025, (b) Fo = 0.001, (c) Fo = 0.002, and (d) Fo = 0.008.

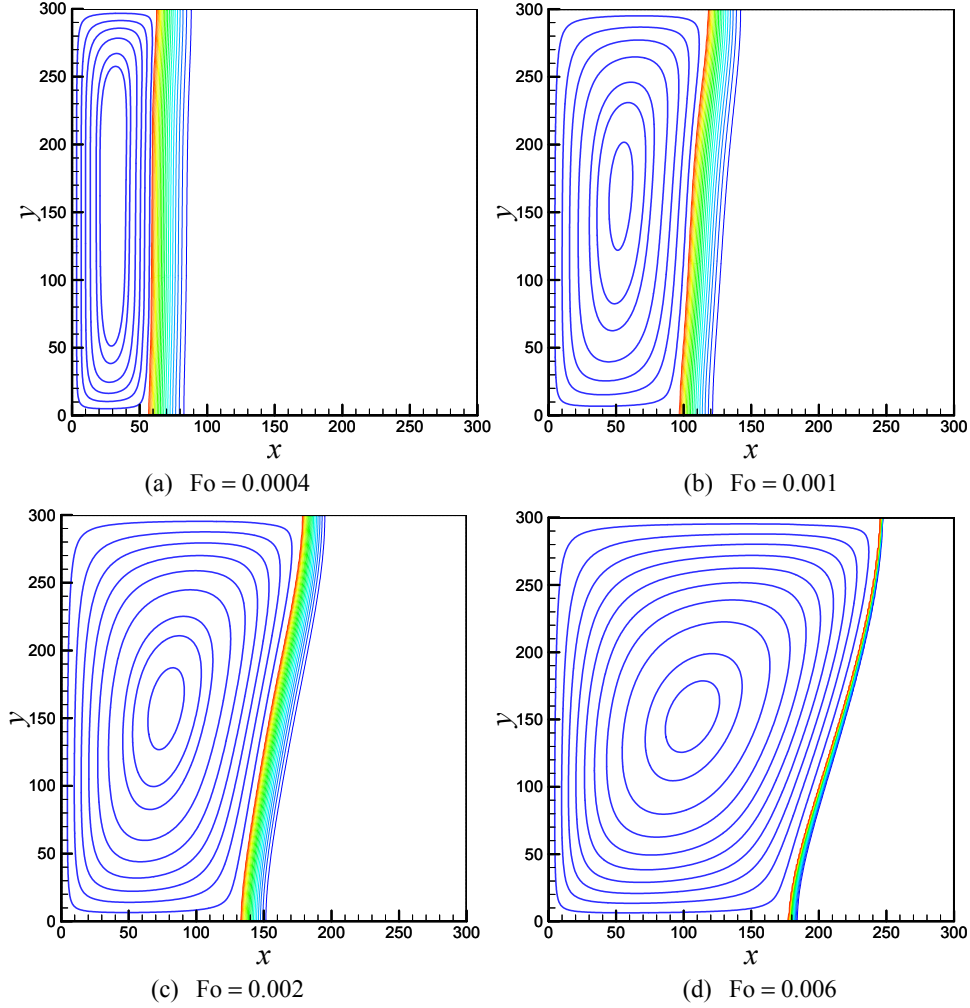

(a) Fo = 0.0004

(b) Fo = 0.001

(c) Fo = 0.002

(d) Fo = 0.006

FIG. 7. Streamlines with the phase interface at different Fourier numbers for $Ra = 10^8$. (a) Fo = 0.0004, (b) Fo = 0.001, (c) Fo = 0.002, and (d) Fo = 0.006.

The temperature profiles at the mid-height ($y/L = 0.5$) of the cavity at different Fourier numbers for $Ra = 10^6$ and $10^8$ are shown in Fig. 8. As can be seen in the figure, the temperature profiles of the PCM and metal foam develop together in a coupled manner. Initially (Fo = 0.00005), the metal foam-to-PCM temperature difference is rather high, but it progressively decreases with the Fourier number. At Fo = 0.006, the temperature profiles of the PCM and metal foam are seen to be nearly identical, which indicates that the thermal non-equilibrium effect between the PCM and metal foam is weak. Fig. 8 clearly shows that the maximum metal foam-to-PCM temperature difference appears near the phase interface. For comparison, the FVM results [15] are also presented in the figure (for clarity, the FVM results at $Fo = 0.006$ are not presented). It can be observed from the figure that the present results agree well with the FVM results reported in the literature. The variations of the total liquid fraction with the Fourier number for $Ra = 10^6$ are shown in Fig. 9. As shown in the figure, the metal foam helps utilize the PCM much more effectively.

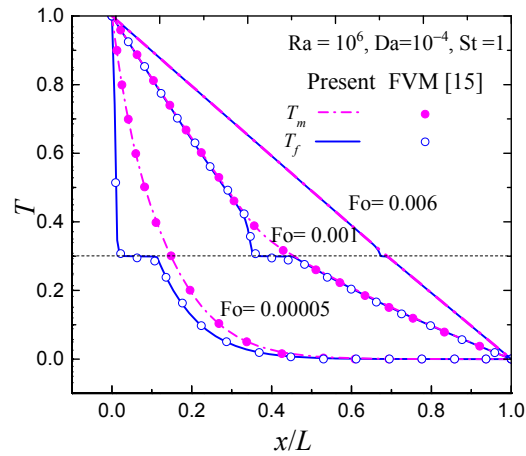

(a) $Ra = 10^6$

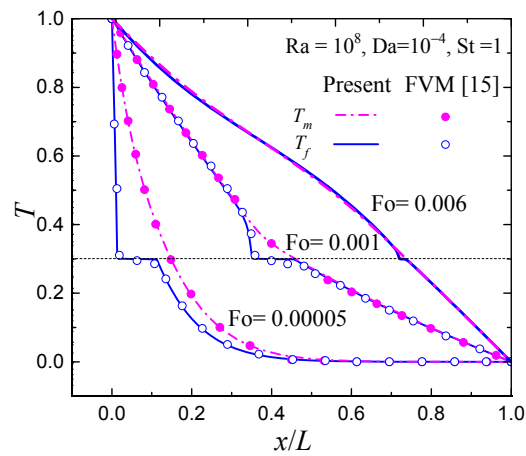

(b) $Ra = 10^8$

FIG. 8. Temperature profiles at the mid-height ($y/L = 0.5$) of the cavity for $Ra = 10^6$ (a) and

$Ra = 10^8$ (b) at different Fourier numbers. For clarity, the FVM results [15] at Fo = 0.006 are not presented.

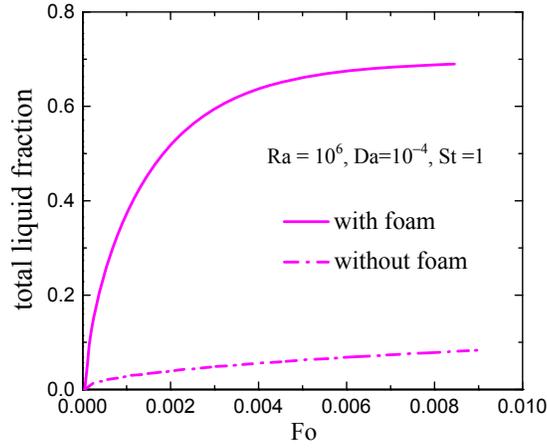

FIG. 9. The variations of the total liquid fraction with the Fourier number for $Ra = 10^6$.

## V. COMPARISONS AND DISCUSSIONS

In Section IV, the accuracy and effectiveness of the enthalpy-based MRT-LB method have been demonstrated. For melting coupled with natural convection in metal-foam-based PCMs, the fluid flow and heat transfer processes during solid-liquid phase change are rather complicated. In this section, comparisons and discussions are made to offer some insights into the roles of the collision model, volumetric LB scheme, enthalpy formulation, and relaxation rate $\zeta_e$ in the present method. Unless otherwise specified, all the simulations are carried out with identical initial and boundary conditions at a fixed Rayleigh number $Ra = 10^8$, and the other parameters can be found in Section IV B.

### A. MRT vs. BGK

The advantage of MRT collision model over BGK collision model is shown in this subsection. In the present study, the BGK results denote that the temperature fields are solved by BGK-LB models [$\zeta_i = 1/\tau_g$, $\eta_i = 1/\tau_h$, and the equilibrium distributions are given by Eqs. (40) and (49)], while the flow field is still solved by the MRT-LB model presented in Section III A. In Fig. 10, the liquid fraction distributions obtained by BGK and MRT collision models at different Fourier numbers are shown. It is very clear that the phase interface obtained by BGK exhibits significant oscillations [see Fig. 10(a)], which is predominantly due to the numerical diffusion across phase interface. On the contrary, the numerical diffusion across phase interface is almost invisible in the MRT results [see Fig. 10(b)]. With additional degrees of freedom, the MRT collision model has the ability to reduce the numerical diffusion across phase interface. In addition, it should be noted that the step-like behavior of the liquid fraction distribution near the liquid/mushy interface [see Fig. 10(b)] is caused by the inevitable numerical error due to the enthalpy formulation.

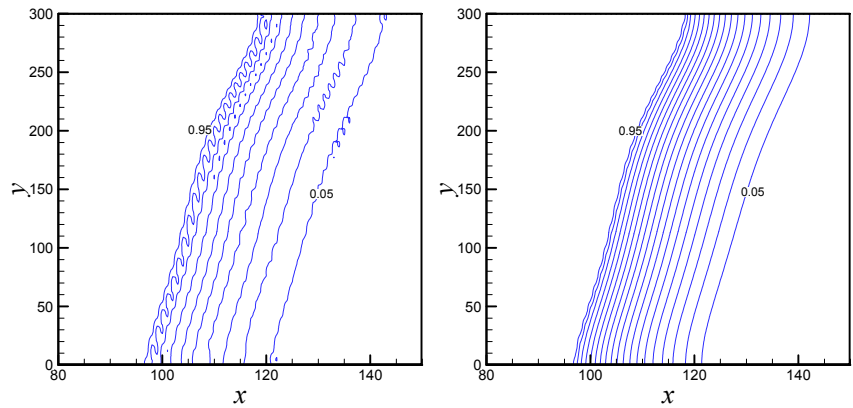

(a)  Fo = 0.001

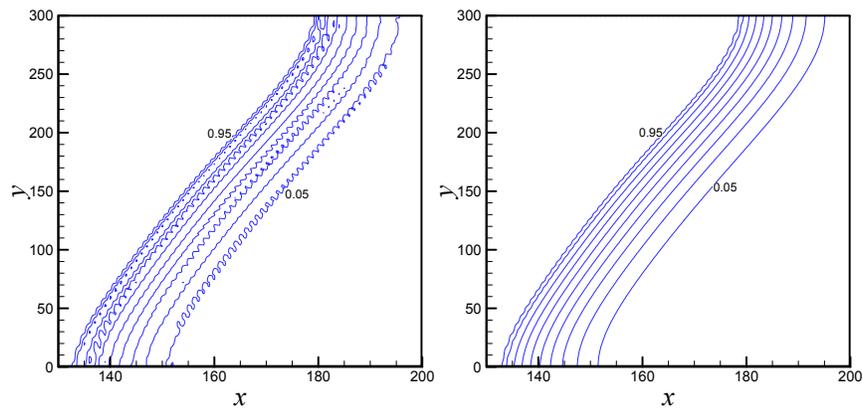

(b)  Fo = 0.002

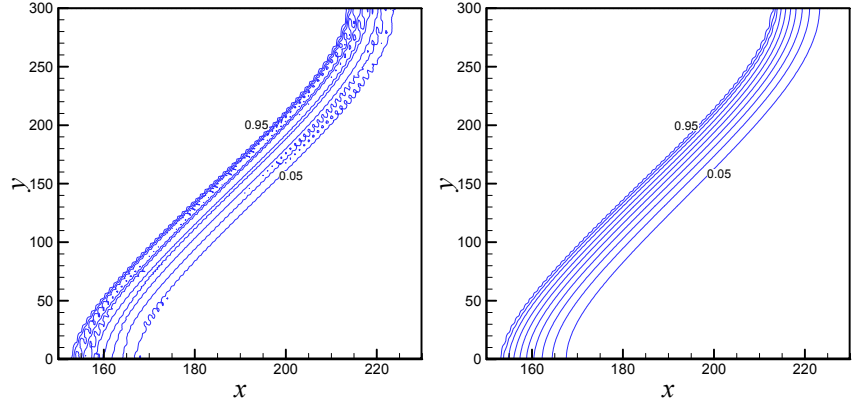

(c) Fo = 0.003

FIG. 10. Local enlargement view of the liquid fraction distributions obtained by BGK (left) and MRT (right) collision models at different Fourier numbers. (a) Fo = 0.001, (b) Fo = 0.002, and (c) Fo = 0.003.

### B. Volumetric LB scheme vs. bounce-back scheme

In the literature [62], the bounce-back scheme was used to impose the no-slip velocity condition on the phase interface and in the solid phase region. Although this approach has some drawbacks [see Ref. [67] for details], it can produce reasonable results when the phase interface occupies one or two lattices. However, for solid-liquid phase change heat transfer in metal foams under LTNE condition, it is not appropriate to use the bounce-back scheme to impose the no-slip velocity condition because the phase interface is actually a region (the so-called interface region or mushy zone) with a certain thickness during phase change process (see Figs. 6 and 7). In what follows, comparisons between the volumetric LB scheme and bounce-back scheme are made to demonstrate this point. In Fig. 11, the streamlines at different Fourier numbers are shown. Clearly, significant small-scale (of the order of lattice size) oscillations can be seen in the streamlines obtained by bounce-back scheme [see Fig. 11(a)], while the streamlines obtained by volumetric LB scheme are smooth [see Fig. 11(b)].

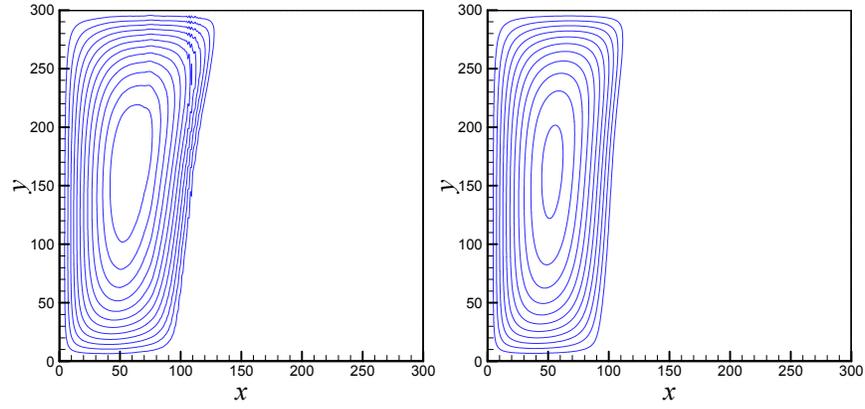

(a) Fo = 0.001

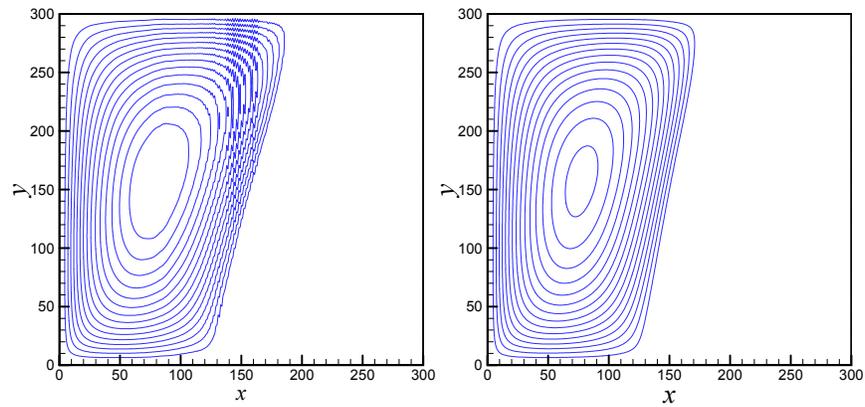

(b) Fo = 0.002

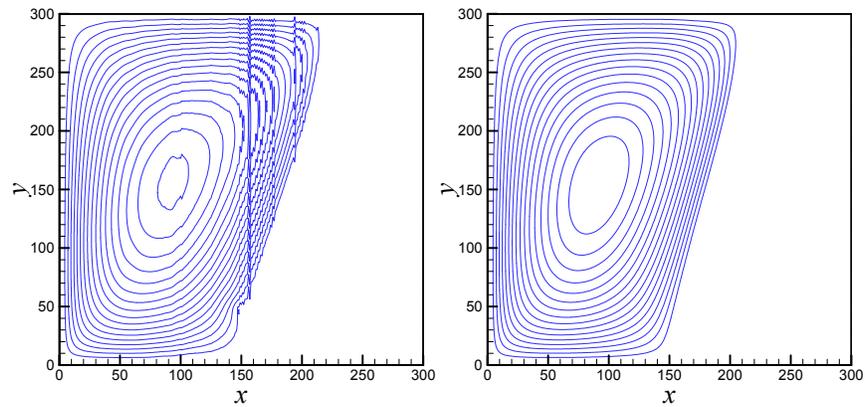

(c) Fo = 0.003

FIG. 11. The streamlines obtained by bounce-back scheme (left) and volumetric LB scheme (right) at different Fourier numbers. (a) Fo = 0.001, (b) Fo = 0.002, and (c) Fo = 0.003.

Fig. 12 shows the local enlargement view of the flow fields in the vicinity of interface region obtained by bounce-back scheme and volumetric LB scheme at $Fo = 0.002$. In the flow field obtained by bounce-back scheme, nonphysical oscillations occur near the phase interface [marked by the red circles in Fig. 12(a), and note that, $f_l = 0.5$ is defined as the phase interface]. On the contrary, the flow field obtained by volumetric LB scheme [see Fig. 12(b)] is rather reasonable. As can be seen in Fig. 12(b), the flow in the interface region is much weaker than that in the liquid phase region near the liquid/mushy interface. It is found that $u_y$ in the liquid phase region near the liquid/mushy interface is of order $O(10^{-3})$, while in the interface region where $f_l < 0.5$, $u_y$ is of order $O(10^{-5})$ or less. In the solid phase region ($f_l = 0$), the velocity is zero ($\mathbf{u} = 0$) at each lattice node. Obviously, the flow in the interface region ($0.5 < f_l < 1$) is rather different from that in the liquid phase region. For the phase interface, the following phenomena can be observed. First, the phase interface obtained by bounce-back scheme exhibits significant oscillations. Second, as compared with the volumetric LB scheme result, the phase interface obtained by bounce-back scheme moves faster near the top wall, but slower near the bottom wall (see top-left and top-right of Fig. 12). From the above comparisons, it can be concluded that the bounce-back scheme is not suitable for imposing the no-slip velocity condition in the interface region, while the volumetric LB scheme [67] is recommended.

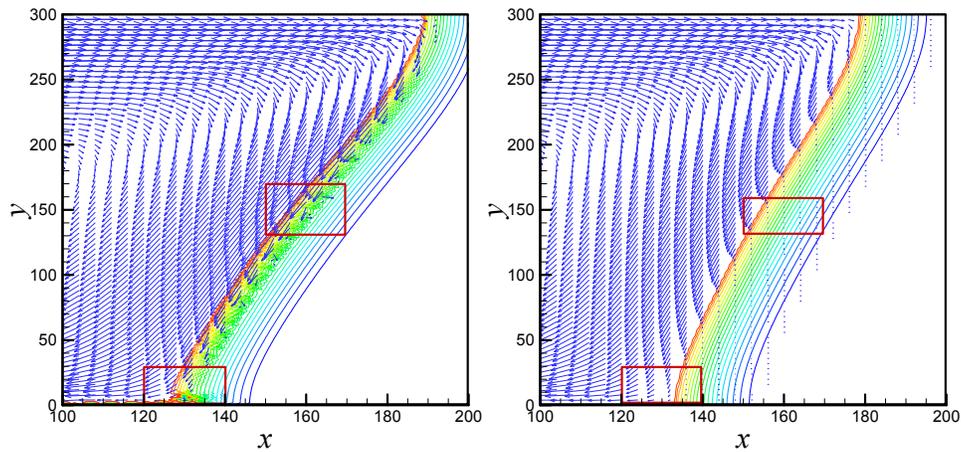

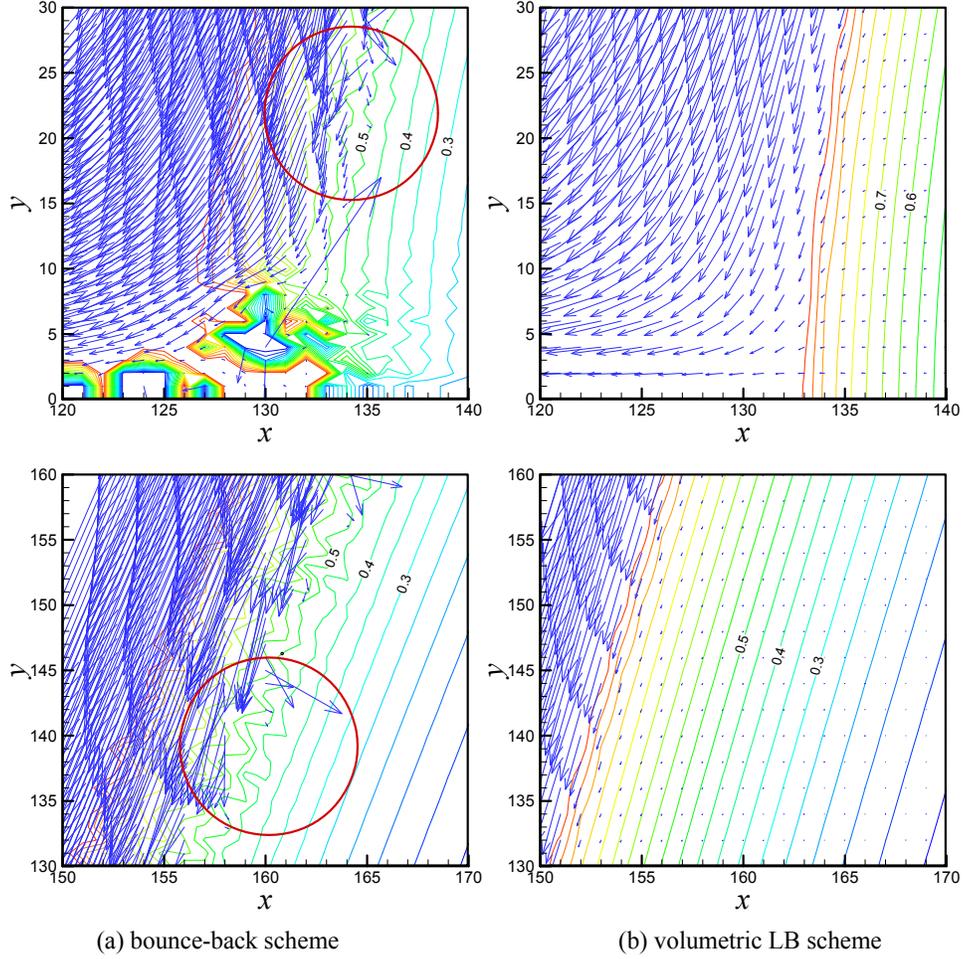

(a) bounce-back scheme  (b) volumetric LB scheme

FIG. 12. Local enlargement view of the flow fields in the vicinity of interface region obtained by bounce-back scheme (a) and volumetric LB scheme (b) at Fo = 0.002.

**C. Present enthalpy scheme vs. iteration enthalpy scheme**

In previous studies [21-23], the nonlinear latent heat source term [the underlined term in Eq. (5)] is treated as a source term in the LB equation of the PCM temperature field, which makes the explicit time-matching LB equation to be implicit. Therefore, the iteration enthalpy scheme [60] is needed so as to obtain the convergent solution of the implicit LB equation. By using the present enthalpy scheme, the iteration procedure can be avoided in simulations. In Table I, we compare the CPU time of the present enthalpy scheme with that of the iteration enthalpy scheme at Fo = 0.002. The simulations are performed on a computer with a quad-core 2.33 GHz processor. It can be seen that the CPU time of the present enthalpy scheme is about one-sixth of that of the iteration enthalpy scheme. Without the iteration procedure, the present method has much higher computational efficiency as compared with previous studies [21-23].

TABLE I. Comparison of the CPU time of the present enthalpy scheme with that of the iteration enthalpy scheme at $\text{Fo} = 0.002$.

| Grid size ( Ra ) | Steps | Method | CPU time (s) | Steps/CPU time |
|---|---|---|---|---|
| $150 \times 150$ ($10^6$) | $9 \times 10^4$ | Present | 423.79 | 212.37 |
| | | Iteration | 2725.45 | 33.02 |
| $300 \times 300$ ($10^8$) | $3.6 \times 10^5$ | Present | 11611.81 | 31.0 |
| | | Iteration | 74043.11 | 4.86 |

In Fig. 13, the PCM temperature profiles at the mid-height of the cavity obtained by the present enthalpy scheme and the iteration enthalpy scheme for $\text{Ra} = 10^6$ at $\text{Fo} = 0.002$ are presented. In simulations, we set $T_{\text{melt}} = 0$ and $H_v = 0$. It can be seen that the temperature obtained by the iteration enthalpy scheme in the solid phase (near the phase interface) is a little lower than $T_{\text{melt}}$ (the minimum temperature is $-9.72 \times 10^{-3}$), which is caused by the negative numerical diffusion across phase interface. On the contrary, the temperature obtained by the present enthalpy scheme in the solid phase can precisely keep at $T_{\text{melt}}$, which indicates that the numerical diffusion across phase interface can be significantly reduced by the present method. The above comparisons confirm that the present method is superior over the iteration method in terms of accuracy and computational efficiency.

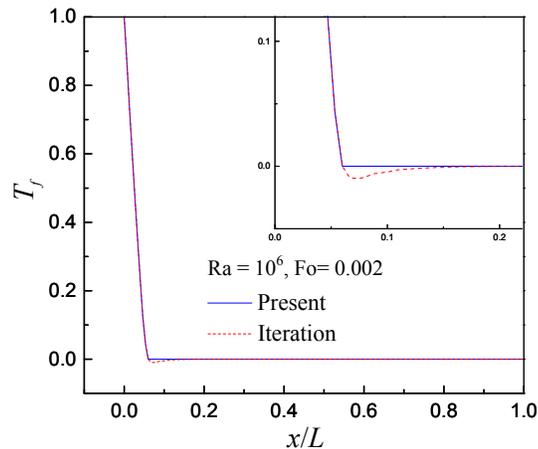

FIG. 13. The PCM temperature profiles at the mid-height of the cavity obtained by the present enthalpy scheme and the iteration enthalpy scheme for $Ra = 10^6$ at $Fo = 0.002$

**D. The effect of the relaxation rate $\zeta_e$**

As shown in Section V A, the numerical diffusion across phase interface can be considerably reduced by the MRT collision model with $\zeta_e = 2 - \zeta_\alpha$. It should be noted that the relaxation rate $\zeta_e$ has apparent influence on the phase interface. To confirm this statement, numerical simulations are carried out for different values of $\zeta_e$. In Fig. 14, the local enlargement view of the phase interfaces for different values of $\zeta_e$ at $Fo = 0.002$ are presented. Form the figure it can be observed that significant oscillations appear at $\zeta_e = 1.8$. As $\zeta_e$ decreases to 0.1, the numerical diffusion is not apparent, and the phase interface is similar to that at $\zeta_e = 2 - \zeta_\alpha$. Actually, with the given parameters (see Section IV B), $\zeta_e$ equals to 0.0296 when it is determined by $\zeta_e = 2 - \zeta_\alpha$. To reduce the numerical diffusion across phase interface, it is recommended that $\zeta_e = 2 - \zeta_\alpha$. However, solid-liquid phase change heat transfer in metal foams is much more complicated than that in the absence of a porous medium, further study about the effect of the relaxation rate $\zeta_e$ is still needed.

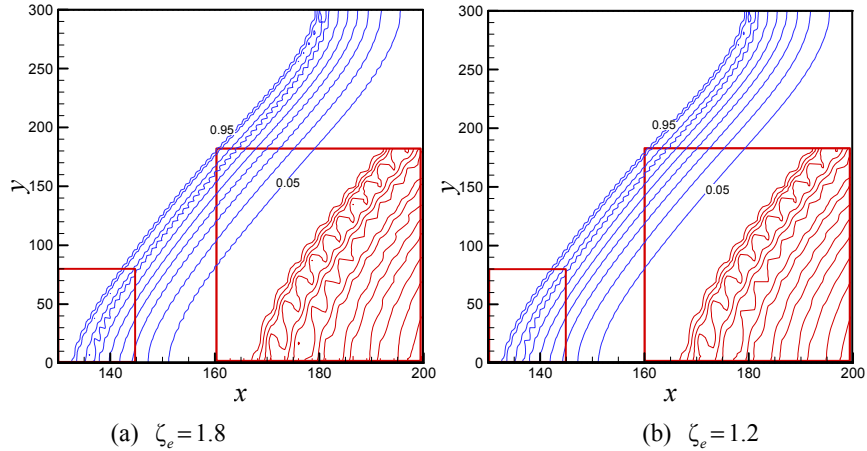

(a) $\zeta_e = 1.8$     (b) $\zeta_e = 1.2$

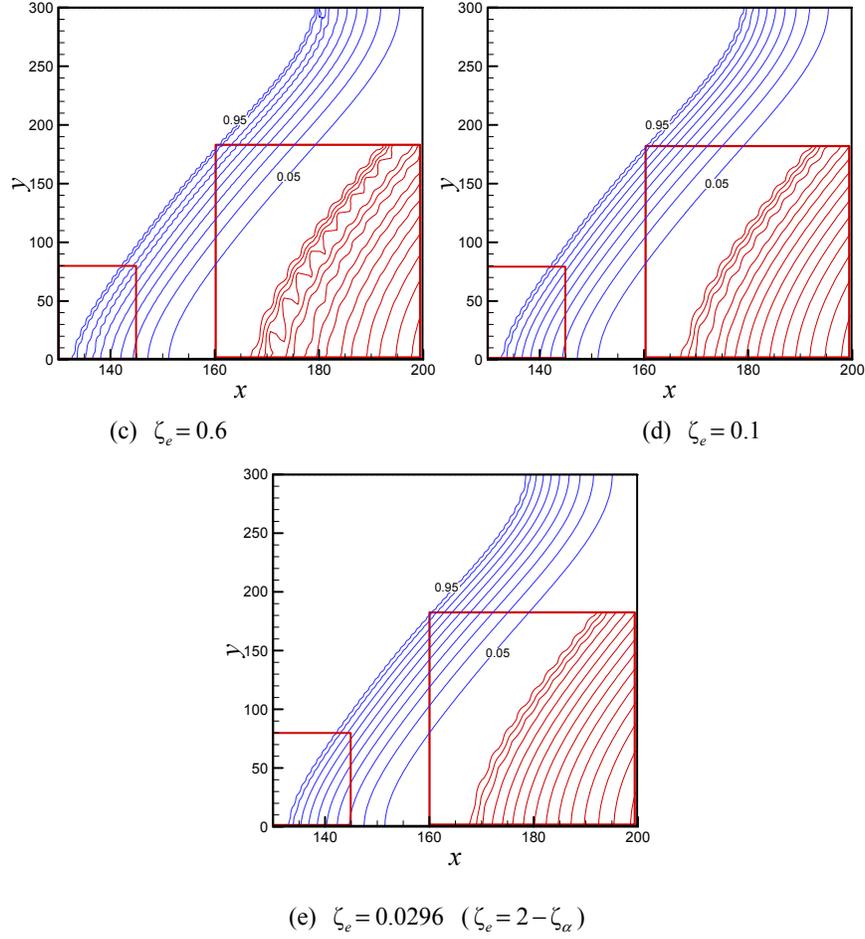

(c) $\zeta_e = 0.6$      (d) $\zeta_e = 0.1$

(e) $\zeta_e = 0.0296$   ($\zeta_e = 2 - \zeta_\alpha$)

FIG. 14. Local enlargement view of the phase interfaces for different values of $\zeta_e$ at Fo = 0.002.

## VI. CONCLUSIONS

In summary, an enthalpy-based MRT-LB method has been developed for solid-liquid phase change heat transfer in metal foams under LTNE condition. In the method, the moving solid-liquid phase interface is implicitly tracked through the liquid fraction, which is simultaneously obtained when the energy equations of PCM and metal foam are solved. The present method has three distinctive features. First, the iteration procedure has been avoided, thus it retains the inherent merits of the standard LB method and is superior over the iteration method in terms of accuracy and computational efficiency. Second, by using the volumetric LB scheme, the no-slip velocity condition in the interface and solid phase regions can be accurately realized. Moreover, the MRT collision model is employed, and with additional degrees of freedom, it has the ability to reduce the numerical diffusion across phase interface induced by solid-liquid phase change. For solid-liquid phase change heat transfer in metal foams, it has been unequivocally demonstrated that the MRT method is superior over its BGK counterpart in terms of accuracy and numerical stability.

Detailed numerical tests of the enthalpy-based MRT-LB method are carried out for two types of solid-liquid phase change heat transfer problems, including the conduction-induced solidification in a semi-infinite domain and melting coupled with natural convection in a square cavity filled with metal-foam-based PCM. It is found that the present results are in good agreement with the FDM or FVM results, which demonstrate that the present method can be served as an accurate and efficient numerical tool for studying metal foam enhanced solid-liquid phase change heat transfer in LHS. Finally, comparisons and discussions are made to offer some insights into the roles of the collision model, volumetric LB scheme, enthalpy formulation, and relaxation rate $\zeta_e$ in the enthalpy-based MRT-LB method, which are very useful for practical applications.

## ACKNOWLEDGEMENTS

This work was financially supported by the Major Program of the National Natural Science Foundation of China (Grant No. 51590902) and the National Key Basic Research Program of China (973 Program) (2013CB228304).

**APPENDIX A: Chapman-Enskog analysis of the MRT-LB model for PCM temperature field**

In this Appendix, the Chapman-Enskog analysis [76] is employed to derive the macroscopic energy equation of the MRT-LB equation (31). To this end, the following multiscale expansions of $\mathbf{n}_g$, the derivatives of time and space, and the source term are introduced

$$\mathbf{n}_g = \mathbf{n}_g^{(0)} + \epsilon \mathbf{n}_g^{(1)} + \epsilon^2 \mathbf{n}_g^{(2)} + \cdots, \quad \partial_t = \epsilon \partial_{t_1} + \epsilon^2 \partial_{t_2}, \quad \nabla = \epsilon \nabla_1, \quad Sr_f = \epsilon Sr_f^{(1)}, \tag{A1}$$

where $\epsilon$ ($\epsilon = \delta_t$) is a small expansion parameter. Taking a second-order Taylor series expansion to Eq. (31), we can obtain

$$(\mathbf{I}\partial_t + \mathbf{E} \cdot \nabla)\mathbf{n}_g + \frac{\delta_t}{2}(\mathbf{I}\partial_t + \mathbf{E} \cdot \nabla)^2 \mathbf{n}_g = -\frac{\Theta}{\delta_t}(\mathbf{n}_g - \mathbf{n}_g^{eq}) + \hat{\mathbf{S}}_{PCM} + \tilde{\mathbf{S}}_{PCM} + O(\delta_t^2), \tag{A2}$$

where $\mathbf{E} = (\mathbf{E}_x, \mathbf{E}_y)^T$, in which $\mathbf{E}_\beta = \mathbf{M}[\text{diag}(e_{0\beta}, e_{1\beta}, \ldots, e_{8\beta})]\mathbf{M}^{-1}$ ($\beta = x, y$), and

$$\hat{\mathbf{S}}_{PCM} = Sr_f(1, -2, 1, 0, 0, 0, 0, 0, 0)^T, \tag{A3}$$

$$\tilde{\mathbf{S}}_{PCM} = \frac{1}{2}\delta_t \partial_t Sr_f(1, -2, 1, 0, 0, 0, 0, 0, 0)^T. \tag{A4}$$

Using the multiscale expansions given by Eq. (A1), the following equations in the consecutive orders of $\epsilon$ in moment space can be obtained

$$\epsilon^0: \quad \mathbf{n}_g^{(0)} = \mathbf{n}_g^{eq}, \tag{A5}$$

$$\epsilon^1: \quad (\mathbf{I}\partial_{t_1} + \mathbf{E} \cdot \nabla_1)\mathbf{n}_g^{(0)} = -\frac{\Theta}{\delta_t}\mathbf{n}_g^{(1)} + \hat{\mathbf{S}}_{PCM}, \tag{A6}$$

$$\epsilon^2: \quad \partial_{t_2}\mathbf{n}_g^{(0)} + (\mathbf{I}\partial_{t_1} + \mathbf{E} \cdot \nabla_1)\mathbf{n}_g^{(1)} + \frac{\delta_t}{2}(\mathbf{I}\partial_{t_1} + \mathbf{E} \cdot \nabla_1)^2 \mathbf{n}_g^{(0)} = -\frac{\Theta}{\delta_t}\mathbf{n}_g^{(2)} + \tilde{\mathbf{S}}_{PCM}, \tag{A7}$$

Using Eq. (A6), Eq. (A7) can be rewritten as

$$\epsilon^2: \quad \partial_{t_2}\mathbf{n}_g^{(0)} + (\mathbf{I}\partial_{t_1} + \mathbf{E} \cdot \nabla_1)\left(\mathbf{I} - \frac{\Theta}{2}\right)\mathbf{n}_g^{(1)} = -\frac{\Theta}{\delta_t}\mathbf{n}_g^{(2)} - \frac{\delta_t}{2}\mathbf{E} \cdot \nabla_1 \hat{\mathbf{S}}_{PCM}. \tag{A8}$$

Writing out the equations for the conserved moment $n_{g0}$ ($n_{g0} = H_f$) of Eqs. (A5), (A6) and (A8), we can obtain

$$\epsilon^0: \quad n_{g0}^{(0)} = n_{g0}^{eq}, \tag{A9}$$

$$\epsilon^1: \quad \partial_{t_1} n_{g0}^{(0)} + c\left[\partial_{x1} n_{g3}^{(0)} + \partial_{y1} n_{g5}^{(0)}\right] = -\frac{\zeta_0}{\delta_t} n_{g0}^{(1)} + Sr_f, \tag{A10}$$

$$\epsilon^2: \quad \partial_{t_2} n_{g0}^{(0)} + \partial_{t_1}\left[\left(1 - \frac{\zeta_0}{2}\right)n_{g0}^{(1)}\right] + c\nabla_1 \cdot \left[\begin{pmatrix} 1-\zeta_3/2 & 0 \\ 0 & 1-\zeta_5/2 \end{pmatrix}\begin{pmatrix} n_{g3}^{(1)} \\ n_{g5}^{(1)} \end{pmatrix}\right] = -\frac{\zeta_0}{\delta_t} n_{g0}^{(2)}, \tag{A11}$$

According to Eq. (A9), we have

$$n_{g0}^{(k)} = 0, \quad \forall k \geq 1. \tag{A12}$$

With the aid of Eqs. (A9) and (A12), we can obtain

$$\epsilon^1: \ \partial_{t_1} H_f + \nabla_1 \cdot \left( \frac{c_{pl} T_f \mathbf{u}}{\phi} \right) = Sr_f, \tag{A13}$$

$$\epsilon^2: \ \partial_{t_2} H_f + c\nabla_1 \cdot \left[ \begin{pmatrix} 1-\zeta_3/2 & 0 \\ 0 & 1-\zeta_5/2 \end{pmatrix} \begin{pmatrix} n_{g3}^{(1)} \\ n_{g5}^{(1)} \end{pmatrix} \right] = 0. \tag{A14}$$

According to Eq. (A6), we have

$$-\frac{\zeta_3}{\delta_t} n_{g3}^{(1)} = \partial_{t_1} n_{g3}^{(0)} + c \left[ \partial_{x1} \left( \tfrac{2}{3} n_{g0}^{(0)} + \tfrac{1}{6} n_{g1}^{(0)} + \tfrac{1}{2} n_{g7}^{(0)} \right) + \partial_{y1} n_{g8}^{(0)} \right]$$

$$= \frac{1}{c} \partial_{t_1} \left( \frac{c_{pl} T_f u_x}{\phi} \right) + c \partial_{x1} \left( \frac{1}{3} c_{f,\mathrm{ref}} T_f \right), \tag{A15}$$

$$-\frac{\zeta_5}{\delta_t} n_{g5}^{(1)} = \partial_{t_1} n_{g5}^{(0)} + c \left[ \partial_{x1} n_{g8}^{(0)} + \partial_{y1} \left( \tfrac{2}{3} n_{g0}^{(0)} + \tfrac{1}{6} n_{g1}^{(0)} - \tfrac{1}{2} n_{g7}^{(0)} \right) \right]$$

$$= \frac{1}{c} \partial_{t_1} \left( \frac{c_{pl} T_f u_y}{\phi} \right) + c \partial_{y1} \left( \frac{1}{3} c_{f,\mathrm{ref}} T_f \right). \tag{A16}$$

Substituting Eqs. (A15) and (A16) into Eq. (A14), the following equation can be obtained

$$\epsilon^2: \ \partial_{t_2} H_f = \nabla_1 \cdot \left\{ \delta_t \begin{pmatrix} \zeta_3^{-1} - \tfrac{1}{2} & 0 \\ 0 & \zeta_5^{-1} - \tfrac{1}{2} \end{pmatrix} \left[ \partial_{t_1} \left( \frac{c_{pl} T_f \mathbf{u}}{\phi} \right) + \frac{c^2}{3} \nabla_1 \left( c_{f,\mathrm{ref}} T_f \right) \right] \right\}. \tag{A17}$$

Note that $\nabla_1 \left( c_{f,\mathrm{ref}} T_f \right) = c_{f,\mathrm{ref}} \nabla_1 T_f$, combining Eqs. (A13) and (A17) leads to the following macroscopic energy equation

$$\frac{\partial H_f}{\partial t} + \nabla \cdot \left( \frac{c_{pl} T_f \mathbf{u}}{\phi} \right) = \nabla \cdot \left[ \alpha_{f,\mathrm{ref}} c_{f,\mathrm{ref}} \nabla T_f + \delta_t \left( \zeta_\alpha^{-1} - 0.5 \right) \epsilon \partial_{t_1} \left( \frac{c_{pl} T_f \mathbf{u}}{\phi} \right) \right] + Sr_f, \tag{A18}$$

where $\alpha_{f,\mathrm{ref}}$ is the reference thermal diffusivity

$$\alpha_{f,\mathrm{ref}} = \frac{k_f}{\rho_l c_{f,\mathrm{ref}}} = c_s^2 \left( \zeta_\alpha^{-1} - \frac{1}{2} \right) \delta_t. \tag{A19}$$

**APPENDIX B: MRT-LB MODELS FOR TEMPERATURE FIELDS BASED ON D2Q5**

**LATTICE**

The MRT-LB models for the temperature fields based on D2Q5 lattice are briefly presented. The five discrete velocities $\{\mathbf{e}_i\}$ of the D2Q5 lattice are given by Eq. (11). For the D2Q5 model, the transformation matrix is given by [78,79]

$$\mathbf{M} = \begin{bmatrix} 1 & 1 & 1 & 1 & 1 \\ 0 & 1 & 0 & -1 & 0 \\ 0 & 0 & 1 & 0 & -1 \\ 0 & 1 & 1 & 1 & 1 \\ 0 & 1 & -1 & 1 & -1 \end{bmatrix}. \tag{B1}$$

For the enthalpy-based MRT-LB model, the equilibrium moment $\mathbf{n}_g^{eq}$ can be chosen as

$$\mathbf{n}_g^{eq} = \left( H_f, \frac{c_{pl}T_f u_x}{\phi c}, \frac{c_{pl}T_f u_y}{\phi c}, \varpi c_{f,\text{ref}} T_f, 0 \right)^{\mathrm{T}}, \tag{B2}$$

where $\varpi \in (0,1)$. The source term in moment space is given by $\mathbf{S}_{\text{PCM}} = S_{\text{PCM}} (1, 0, 0, \varpi, 0)^{\mathrm{T}}$, the relaxation matrix is given by $\mathbf{\Theta} = \text{diag}(\zeta_0, \zeta_\alpha, \zeta_\alpha, \zeta_e, \zeta_\varepsilon)$, and the reference thermal diffusivity is defined as $\alpha_{f,\text{ref}} = c_{sT}^2 (\zeta_\alpha^{-1} - 0.5) \delta_t$. To reduce the numerical diffusion across the phase interface, $\zeta_e$ is determined by $\zeta_e = 2 - \zeta_\alpha$. The equilibrium enthalpy distribution function $g_i^{eq}$ in velocity space is

$$g_i^{eq} = \begin{cases} H_f - \varpi c_{f,\text{ref}} T_f, & i = 0, \\ \dfrac{1}{4} \varpi c_{pl} T_f \left( \dfrac{c_{f,\text{ref}}}{c_{pl}} + \dfrac{\mathbf{e}_i \cdot \mathbf{u}}{\phi c_{sT}^2} \right), & i = 1 \sim 4, \end{cases} \tag{B3}$$

where $c_{sT}^2 = c^2 \varpi / 2$ ($c_{sT}$ is the sound speed of the D2Q5 model).

For the internal-energy-based MRT-LB model, the equilibrium moment $\mathbf{n}_h^{eq}$ can be chosen as

$$\mathbf{n}_h^{eq} = \left( c_{pm} T_m, 0, 0, \varpi c_{m,\text{ref}} T_m, 0 \right)^{\mathrm{T}}. \tag{B4}$$

The source term in moment space is given by $\mathbf{S}_{\text{metal}} = S_{\text{metal}} (1, 0, 0, \varpi, 0)^{\mathrm{T}}$, the relaxation matrix is given by $\mathbf{Q} = \text{diag}(\eta_0, \eta_\alpha, \eta_\alpha, \eta_e, \eta_\varepsilon)$, the reference thermal diffusivity is defined as $\alpha_{m,\text{ref}} = c_{sT}^2 (\eta_\alpha^{-1} - 0.5) \delta_t$, and the equilibrium internal-energy distribution function $h_i^{eq}$ in velocity space is

$$h_i^{eq} = \begin{cases} c_{pm} T_m - \varpi c_{m,\text{ref}} T_m, & i = 0, \\ \dfrac{1}{4} \varpi c_{m,\text{ref}} T_m, & i = 1 \sim 4. \end{cases} \tag{B5}$$

# REFERENCES

[1] B. Zalba, J.M. Marín, L.F. Cabeza, and H. Mehling, Appl. Therm. Eng. **23**, 251 (2003).

[2] S.V. Garimella, Microelectron. J. **37**, 1165 (2006).

[3] A. Sharma, V.V. Tyagi, C.R. Chen, and D. Buddhi, Renew. Sustain. Energy Rev. **13**, 318 (2009).

[4] F. Agyenim, N. Hewitt, P. Eames, and M. Smyth, Renew. Sustain. Energy Rev. **14**, 615 (2010).

[5] Y.B. Tao and Y.L. He, Appl. Energy **88**, 4172 (2011).

[6] L. Fan and J.M. Khodadadi, Renew. Sustain. Energy Rev. **15**, 24 (2011).

[7] C.Y. Zhao, Int. J. Heat Mass Transfer **55**, 3618 (2012).

[8] E.A. Ellinger and C. Beckermann, Exp. Therm. Fluid Sci. **4**, 619 (1991).

[9] S.T. Hong and D.R. Herling, Adv. Eng. Mater. **9**, 554 (2007).

[10] C.Y. Zhao and Z.G. Wu, Sol. Energy Mater. Sol. Cells **95**, 636 (2011).

[11] Z.G. Wu and C.Y. Zhao, Sol. Energy **85**, 1371 (2011).

[12] S. Mancin, A. Diani, L. Doretti, K. Hooman, and L. Rossetto, Int. J. Therm. Sci. **90**, 79 (2015).

[13] X. Tong, J.A. Khan, and M. RuhulAmin, Numer. Heat Transfer A **30**, 125 (1996).

[14] O. Mesalhy, K. Lafdi, A. Elgafy, and K. Bowman, Energy Convers. Manage. **46**, 847 (2005).

[15] S. Krishnan, J.Y. Murthy, and S.V. Garimella, J. Heat Transfer **127**, 995 (2005).

[16] Z. Yang and S.V. Garimella, J. Heat Transfer **132**, 062301 (2010).

[17] Y. Tian and C.Y. Zhao, Energy **36**, 5539 (2011).

[18] W.Q. Li, Z.G. Qu, Y.L. He, and W.Q. Tao, Appl. Therm. Eng. **37**, 1 (2012).

[19] S.S. Sundarram and W. Li, Appl. Therm. Eng. **64**, 147 (2014).

[20] Y. Zhao, C.Y. Zhao, Z.G. Xu, and H.J. Xu, Int. J. Heat Mass Transfer **99**, 170 (2016).

[21] D. Gao, Z. Chen, M. Shi, and Z. Wu, Sci. China Technol. Sci. **53**, 3079 (2010).

[22] D. Gao, Z. Chen, and L. Chen, Int. J. Heat Mass Transfer **70**, 979 (2014).

[23] Y.B. Tao, Y. You, and Y.L. He, Appl. Therm. Eng. **93**, 476 (2016).

[24] G.R. McNamara and G. Zanetti, Phys. Rev. Lett. **61**, 2332 (1988).

[25] F.J. Higuera, S. Succi, and R. Benzi, Europhys. Lett. **9**, 345 (1989).

[26] J.M.V.A. Koelman, Europhys. Lett. **15**, 603 (1991).

[27] Y.H. Qian, D. d'Humières, and P. Lallemand, Europhys. Lett. **17**, 479 (1992).

[28] D. d'Humières, in *Rarefied Gas Dynamics: Theory and Simulations*, Vol. 159 of Progress in Astronautics and Aeronautics, edited by B. D. Shizgal and D. P. Weave (AIAA, Washington, DC,

1992), pp. 450-458.

[29] S. Chen and G.D. Doolen, Annu. Rev. Fluid Mech. **30**, 329 (1998).

[30] S. Succi, *The Lattice Boltzmann Equation for Fluid Dynamics and Beyond* (Clarendon Press, Oxford, 2001).

[31] Y.L. He, Y. Wang, and Q. Li, *Lattice Boltzmann Method: Theory and Applications* (Science Press, Beijing, 2009).

[32] Z.L. Guo and C. Shu, *Lattice Boltzmann Method and its Applications in Engineering* (World Scientific Publishing, Singapore, 2013).

[33] Q. Li, K.H. Luo, Q.J. Kang, Y.L. He, Q. Chen, and Q. Liu, Prog. Energy. Combust. Sci. **52**, 62 (2016).

[34] U. Frisch, B. Hasslacher, and Y. Pomeau, Phys. Rev. Lett. **56**, 1505 (1986).

[35] X. He and L.-S. Luo, Phys. Rev. E **55**, R6333 (1997).

[36] X. He and L.-S. Luo, Phys. Rev. E **56**, 6811 (1997).

[37] F. Sharipov, J. Micromech. Microeng. **9**, 394 (1999).

[38] K. Xu, J. Computat. Phys. **171**, 289 (2001).

[39] H. Liu, K. Xu, T. Zhu, and W. Ye, Comput. Fluids **67**, 115 (2012).

[40] G. Bird, *Molecular gas dynamics and the direct simulation of gas flows* (Clarendon Press, Oxford, 1994).

[41] T. Zhu and W. Ye, Phys. Rev. E **82**, 036308 (2010).

[42] T. Zhu, W. Ye, and J. Zhang, Phys. Rev. E **84**, 056316 (2011).

[43] S. Succi, Eur. Phys. J. B **64**, 471 (2008).

[44] K. Vafai and C.L. Tien, Int. J. Heat Mass Transfer **24**, 195 (1981).

[45] C.T. Hsu and P. Cheng, Int. J. Heat Mass Transfer **33**, 1587 (1990).

[46] P. Nithiarasu, K.N. Seetharamu, and T. Sundararajan, Int. J. Heat Mass Transfer **40**, 3955 (1997).

[47] S. Ergun, Chem. Eng. Prog. **48**, 89 (1952).

[48] K. Vafai, J. Fluid Mech. **147**, 233 (1984).

[49] J.J. Hwang, G.J. Hwang, R.H. Yeh, and C.H. Chao, J. Heat Transfer **124**, 120 (2002).

[50] C. Beckermann and R. Viskanta, Int. J. Heat Mass Transfer **31**, 35 (1988).

[51] G. De Fabritiis, A. Mancini, D. Mansutti, and S. Succi, Int. J. Modern Phys. C **9**, 1405 (1998).

[52] W. Miller, S. Succi, and D. Mansutti, Phys. Rev. Lett. **86**, 3578 (2001).

[53] I. Rasin, W. Miller, and S. Succi, Phys. Rev. E **72**, 066705 (2005).

[54] D. Medvedev and K. Kassner, Phys. Rev. E **72**, 056703 (2005).

[55] W. Miller, I. Rasin, and S. Succi, Physica A **362**, 78 (2006).

[56] D. Medvedev, T. Fischaleck, and K. Kassner, Phys. Rev. E **74**, 031606 (2006).

[57] M. Selzer, M. Jainta, and B. Nestler, Phys. Status Solidi B **246**, 1197-1205 (2009).

[58] D. Medvedev, F. Varnik, and I. Steinbach, Procedia Comput. Sci. **18**, 2512-2520 (2013).

[59] A. Subhedar, I. Steinbach, and F. Varnik, Phys. Rev. E **92**, 023303 (2015).

[60] W.S. Jiaung, J.R. Ho, and C.P. Kuo, Numer. Heat Transfer B **39**, 167 (2001).

[61] S. Chakraborty and D. Chatterjee, J. Fluid Mech. **592**, 155 (2007).

[62] C. Huber, A. Parmigiani, B. Chopard, M. Manga, and O. Bachmann, Int. J. Heat Fluid Flow **29**, 1469 (2008).

[63] D. Gao and Z. Chen, Int. J. Therm. Sci. **50**, 493 (2011).

[64] R. Huang, H. Wu, and P. Cheng, Int. J. Heat Mass Transfer **59**, 295 (2013).

[65] Q. Liu and Y.L. He, Physica A **438**, 94 (2015).

[66] R. Huang and H. Wu, J. Computat. Phys. **294**, 346 (2015).

[67] R. Huang and H. Wu, J. Comput. Phys. **315**, 65 (2016).

[68] Q. Ren and C.L. Chan, Int. J. Heat Mass Transfer **100**, 522 (2016).

[69] R. Huang and H. Wu, J. Comput. Phys. **277**, 305 (2014).

[70] Z. Li, M. Yang, and Y. Zhang, Numer. Heat Transfer B **68**, 1175 (2015).

[71] P. Lallemand and L.-S. Luo, Phys. Rev. E **61**, 6546 (2000).

[72] M.E. McCracken and J. Abraham, Phys. Rev. E **71**, 036701 (2005).

[73] Q. Liu, Y.L. He, Q. Li, and W.Q. Tao, Int. J. Heat Mass Transfer **73**, 761 (2014).

[74] Z. Guo and T.S. Zhao, Phys. Rev. E **66**, 036304 (2002).

[75] B. Shi and Z. Guo, Phys. Rev. E **79**, 016701 (2009).

[76] S. Chapman and T.G. Cowling, *The Mathematical Theory of Non-uniform Gases*, 3rd ed. (Cambridge University Press, Cambridge, 1970).

[77] Z. Guo, C. Zheng, and B. Shi, Phys. Fluids **14**, 2006 (2002).

[78] Q. Liu, Y.L. He, D. Li, and Q. Li, Int. J. Heat Mass Transfer **102**, 1334 (2016).

[79] Q. Liu and Y.L. He, Physica A **465**, 742 (2017).